\documentclass[aps]{revtex4-2}
\usepackage{amsmath,amssymb,bbold}
\usepackage{color,graphicx}
\usepackage{geometry}
\newgeometry{tmargin=2cm, bmargin=3cm, lmargin=1cm, rmargin=1cm}
\newcommand{\be}{\begin{equation}}
\newcommand{\ee}{\end{equation}}
\begin{document}
 \title{L\'{e}vy processes in  bounded domains:  Path-wise
 reflection scenarios   and  signatures of confinement.}
 \author{Piotr Garbaczewski and Mariusz \.{Z}aba}
 \affiliation{Institute of Physics, University of Opole, 45-052 Opole, Poland}
 \date{\today }
 \begin{abstract}
We discuss an impact of  various (path-wise)  reflection-from-the barrier  scenarios   upon confining properties of  a paradigmatic  family   of symmetric  $\alpha $-stable L\'{e}vy processes,  whose permanent residence  in  a  finite  interval   on a line  is secured   by   a two-sided reflection. Depending on the specific reflection "mechanism", the inferred jump-type processes differ in their spectral and statistical characteristics, like  e.g. relaxation properties, and  functional  shapes of  invariant  (equilibrium, or asymptotic near-equilibrium)  probability density functions  in the interval.  The  analysis  is carried out in conjunction with  attempts  to  give meaning to the notion of a reflecting L\'{e}vy process, in  terms of the   domain  of   its motion generator, to which an invariant pdf  (actually an eigenfunction) does   belong.
\end{abstract}
 \maketitle

\section{Motivation}

We consider symmetric L\'{e}vy jump-type stochastic  processes, which are confined in  an  interval $[0,b]\subset R, b>0$, with reflecting endpoints (a concept hampered by ambiguities, to be resolved in below).   The corresponding    random dynamics, usually is formalized in terms of stochastic differential equations  with   reflecting boundary conditions (their meaning needs to be  specified with due care).  The  inferred  time evolution of  associated   probability density functions (always defined on the whole of $R$, albeit with  exterior restrictions in $R\setminus (0,b)$),   in the large time asymptotic  is  expected to show symptoms of  convergence  to an   equilibrium  pdf (stationary or steady state function), which    needs to  belong  to the domain of the  properly   "tailored"  fractional  Laplacian, c.f. \cite{gar,kyprianou,daoud}. The latter  is regarded as the random motion generator of the reflected L\'{e}vy process.

As discussed in \cite{gar}, the assignment of a proper motion generator to  the   L\'{e}vy  jump-type stochastic  process in a bounded domain (any sort, there are many options) is not at all obvious, and   path-wise  reflecting  boundary data need to be confronted and reconciled with a variety of admissible boundary data for the nonlocally defined fractional Laplacian. These include both  spatial  domain restrictions  and  boundary constraints of the Dirichlet and Neumann-type, which define the admissible  function space.

 Here we encounter a number of problems: (i) there is no unique definition of the boundary-data-respecting  fractional Laplacian in the interval, (ii) there is   no  unique technical implementation of the  Neumann-type  reflecting  boundary condition, (iii)   a particular path-wise  procedure, telling  how a reflection is  executed  at  the barrier (e.g. detailed  reflecting boundary  conditions for the  stochastic  process) appears not to be an  innocent choice, and may have a serious impact on the functional shape of asymptotic probability densities (we shall pay some  attention to this point in below), (iv) for each reflection-at-the-barrier scenario, an assignment of the proper  motion generator needs to be enabled, and in reverse; albeit  there is no one-to-one correspondence.

 We anticipate the outcome of our subsequent discussion, by emphasizing the impact  of the physics-oriented reasoning in the study of random motion and L\'{e}vy processes in particular.  In case of bounded domains, the choice of the reflection-at-the boundary "mechanism"   appears to be  a major discrimination tool between different options for what is to be  consistently   named a {\it reflected L\'{e}vy process}.

The term "fractional Laplacian" refers to the  nonlocally defined    operator  $(- \Delta )^{\alpha /2}$,  which is  interpreted   as  the  generator of  a symmetric  $\alpha $-stable L\'{e}vy process on $R$,  \cite{gar}.   Its "tailoring" refers to  a  profound  problem   of deducing  the  appropriate reflection-restricted  form    $(- \Delta )^{\alpha /2} \rightarrow  (- \Delta )^{\alpha /2}_{\cal{R}}$, where the subscript $\cal{R}$  indicates that suitable   boundary conditions  are imposed  upon   $(- \Delta )^{\alpha /2}$.   This includes  both spatial domain restrictions and operator   domain restrictions in the form of  functional constraints of the Neumann-type, see \cite{gar}  and a subsequent discussion in Section V.

It is clear that the   prescribed   path-wise reflection scenario  for the jump-type process  is    encoded in  the inferred     fractional    differential equation $\partial_t \rho (x,t)=   - (- \Delta )^{\alpha /2}_{\cal{R}}  \rho (x,t)$,  governing  the time evolution of probability density functions and setting their asymptotic. Different reflection recipes may  result in inequivalent invariant pdfs. This we shall demonstrate in below.

 For comparison we recall the Brownian form $\partial _t \rho = \Delta \rho $, and recall that the ordinary Laplacian is negative-definite.  We   mention that the  standard  reflected Brownian motion is   understood as  a Wiener process in an interval  with reflecting boundaries.  The casual   Neumann condition, which specifies  values of the  derivative of the stationary pdf at the boundaries,   is  fully   compatible with the path-wise  {\it instantaneous} reflection scenario.   On formal grounds the Brownian reflection mechanism  refers to  so-called reflection principle, which  is known  not to be valid  for L\'{e}vy processes.  That, in view of the  nonlocality   of fractional Laplacians  and discontinuities of      jump-type  sample paths. Accordingly, the   terms  "suitable", "appropriate"   and  "reflecting" become  ambiguous in the  L\'{e}vy processes  context.

In the mathematical literature one encounters attempts  to define reflected L\'{e}vy processes by means of Neumann-type constraints   (like e.g. local and  nonlocal notions of the "normal derivative")  imposed on the spatially constrained fractional Laplacian, \cite{bogdan}-\cite{Abatangelo1}. There is no general consensus concerning the  {\it  proper}  generalisation of the   Neumann condition from the Brownian   to the  (reflecting)  L\'{e}vy  framework. The pertinent Neumann-type conditions happen to be  inequivalent, refer to (induce, or alternatively -  result from)  inequivalent  path-wise  reflection scenarios,    and   might  imply  {\it   incompatible}  profiles  of  the inferred   stationary pdfs.  This  in turn needs to be reconciled with the inherent nonlocality of the    fractional Laplacian $(- \Delta )^{\alpha /2}$, which gives rise to its   varied,  inequivalent  domain-restricted versions, \cite{gar,zaba,gar1,gar2}.

 Let us mention that in Ref. \cite{gar}, in the general discussion of L\'{e}vy processes in bounded domains,  a  rough  distinction between the  Dirichlet and  Neumann boundary data has been encoded in  the  notation $(- \Delta )^{\alpha /2}_{\cal{D}}$  and $(- \Delta )^{\alpha /2}_{\cal{N}}$  respectively.

Our   notation $(- \Delta )^{\alpha /2}_{\cal{R}}$,  refers to an  anticipated  existence of  the  family of  inequivalent  "reflecting processes"  in the interval, with  motion generators subject to inequivalent (albeit semantically "reflecting")  constraints, which we loosely abbreviate as Neumann-type boundary data.  These in turn  rely on  the presumed  microscopic  (path-wise)  reflection  scenarios in the vicinity of the barrier  and/or  at the barrier location.     Moreover, in principle one  may  admit  random dynamics with forth and back  jumps, which   overshoot   barriers  at $0$ and $b$ from the interior of the interval, provided  an {\it instantaneous   return} to $(0,b)$  follows.

Here, it is useful to mention a concept of the  {\it  return} processes, c.f. \cite{elliott,elliott1}, thoroughly  analyzed   for the case of  Cauchy noise ($\alpha $-stable with $\alpha =1$). Notwithstanding, the  instantaneous {\it random  return} scenario  actually appears to  underlie  the concept of  nonlocal Neumann boundary  conditions for   fractional Laplacians in a bounded domain, as  introduced in Refs. \cite{DRV,ros,Abatangelo}.

In the above discussion, we have been somewhat freely moving  between  the   path-wise  and fractional Laplacian  implementations of  L\'{e}vy processes, although they  look disparately diverse. Actually, this is not the case, there is a deep  connection  between them. 

 The  $\alpha $-stable random variable  $X$ (and   the corresponding stochastic process $X(t) \equiv X_t$) can be  introduced by means of its  characteristic function
$\phi (p) = E[\exp(ipX)]$, which   is uniquely related to the  corresponding probability density function $\rho (x)$ by the  Fourier transform $\rho (x)= (1/2\pi ) \int_R \phi (p) \exp(-ipx)  dp$.  (Dimensional constants are scaled away.)

For symmetric L\'{e}vy processes on $R$, we adopt the  logarithmic  parametrization  of the characteristic function:
\be
\ln \phi (p) = - \sigma ^{\alpha } |p|^{\alpha }= -\sigma ^{\alpha } F(p)
\ee
where $\alpha \in (0,2]$  and $\sigma > 0$ is  a scale parameter, related to a  full width of $\rho (x)$ at its  half-maximum  (FWHM), \cite{JW,JW1,JW2,kyprianou}.  Since $\alpha $-stable  pdfs  have no finite variance, the  familiar  notion  of  a  "standard deviation" is undefined.  For   the  exemplary  Cauchy case,   $\alpha =1 $  in Eq. (1),  we have    $\rho _{\alpha =1}(x)= \sigma/\pi (x^2 + \sigma ^2)$ and   the FWHM  reads $2\sigma $.

The scale factor $\sigma $ can  be eliminated from the formalism.   Namely, let us assume that the  stable  random variable $X$  has  a probability distribution  $\rho (x)$   fixed by  (1). We  encode  this  assignment  by  the notation $X \sim  S_{\alpha }(\sigma)$, borrowed from Refs. \cite{JW, JW2}. Once we have given   $X \sim  S_{\alpha }(1)$,  then for  the rescaled random variable $Y= \sigma X$  we have  $Y \sim  S_{\alpha }(\sigma)$. Thus, for a given stability index $\alpha $, the probability distribution  $S_{\alpha }(1)$ actually  stands for a reference one.   From now on we  associate the random variable  $X$  exclusively with $S_{\alpha }(1)$,  i.e. we presume   $X\sim S_{\alpha }(1)$.   This allows us to  proceed  with    $E[\exp(ipX]  =  \exp[- F(p)]$, instead of  Eq. (1) proper.

Since any L\'{e}vy process  has the property   that   for all $t\geq 0$,   there holds
\be
E[\exp(ipX_t)]  =  \exp[- t F(p)],
\ee
and  we  have uniquely determined the time-dependence  $\rho (x) \rightarrow \rho (x,t)$  of the reference pdf, and  its $\sigma $-scaled versions. The generator of such dynamics and the related  fractional  Fokker-Planck equation can be deduced as  follows.

The notation $F(p)= |p|^{\alpha }$  of Eq. (1),   sets a direct  link with the fractional semigroup dynamics,  \cite{gar,zaba,gar1}.  To this end we invoke a substitution procedure, which actually amounts to a canonical quantization step, \cite{quant},  (up to the explicit  presence  of  $\hbar$):
 \begin{equation}
  p \rightarrow \hat{p}=-i \nabla  \Longrightarrow   F(p)\rightarrow F(\hat{p})=  (- \Delta )^{\alpha /2}
 \, . \label{quant}
 \end{equation}
Since we refer to the standard Fourier representation, a casual quantum  mechanical operator  notion  $(\hat{x}f)(x)= xf(x)$ is implicit.  The inferred  semigroup operator $\exp [- t F(\hat{p})$ gives rise to the fractional Fokker-Planck equation (no drifts, the fractional Laplacian is the motion generator)
\be
\partial _t \rho (x,t) = - (- \Delta )^{\alpha /2} \rho (x,t),
\ee
 with $\rho _0(x)$ given as the initial $t=1$  datum.

To justify the recipe  (3) one may invoke the   Fourier multiplier representation  of the  fractional Laplacian, \cite{gar}:
\be
{\cal{F}} [(- \Delta )^{\alpha /2} f](k) = |k|^{\alpha } {\cal{F}} [f](k),
\ee
while remembering that  it is $- (-\Delta )^{\alpha /2}$,  which is a fractional analog of the   Laplacian  $\Delta $.

We prefer to give meaning to the quantization procedure (3),  by  employing  the L\'{e}vy-Khinchine formula  for the characteristic exponent $F(p)$  of the $\alpha $-stable random variable. In one spatial  dimension, we  ultimately  deal with a  reduced integral expression, (the Cauchy principal value of the integral is implicit):
\begin{equation}
{F(p) = -  \int_{-\infty }^{+\infty } [exp(ipy) - 1]
\nu (dy)}
\end{equation}
 where $\nu (dy)$ stands for the L\'{e}vy measure. In view of (3), we have defined the action of the semigroup generator $- F(\hat{p})$ on  functions in its domain according to:
\begin{equation}
(-  \Delta )^{\alpha /2} f(x) =  F(\hat{p})f(x) =  - (p.v) \int_R [f(x+y)-f(x)] \nu(dy).
 \end{equation}
We emphasize that a generically singular behavior of the L\'{e}vy measure in the vicinity of zero needs the  (counter)term containing $-f(x)$ for consistency reasons.  In the above formulas, the L\'{e}vy measure reads:
\begin{equation}
    \nu (dy) =   {\frac{\mathcal{A}_{\alpha }}{|y|^{1+\alpha }}}\, dy   =  \left[{\Gamma(1+\alpha )\sin\frac{\pi \alpha }{2}}\right] \,    {\frac{dy}{\pi  |y|^{1+\alpha }}}.
\end{equation}

We are exactly at the point where our  main problem can be properly verbalised.   We  are interested  in  symmetric stable processes which are not  running on the whole  real line $R$, but are restricted to  the interval $[0,b]\subset R$, or  - in more  restrictive form - are bound never to leave an  open set  $(0,b)$.
 Here, another   delicate boundary problem appears, since we need to know whether the process may  at all  approach the  interval  boundaries, \cite{bogdan},  and whether or how  their  "overshooting" may be avoided or  somehow   compensated,   \cite{skorohod,As}.

An issue of killed and  taboo L\'{e}vy processes,  which   are  tightly related to exterior Dirichlet boundary data for the fractional Laplacian, has received an ample coverage both in the mathematical and physics-oriented literature, see e.g. \cite{gar} for a sample of relevant references.  Therefore, we leave that topic  aside.

 To the contrary, the problem of reflected L\'{e}vy processes  and their domain-restricted generators, still  remains somewhat enigmatic, \cite{gar,zaba,gar1,gar2}. Quite apart from  the  on-going    mathematical  discussion  of (i) appropriate domain restrictions for the fractional Laplacian \cite{bogdan}-\cite{daoud}, (ii)   path-wise analysis, mostly based on the Skorohod reflection  scenario on the level of  stochastic differential equations with the L\'{e}vy noise, \cite{pilipenko}-\cite{ievlev}.

 As far as the physics-oriented research is concerned, we  adopt  concrete  reflection scenarios, whose usefulness has been  tested in  two active streamlines. Since  the path-wise strategy  involves  Monte Carlo computations,  one can  directly  verify the dependence of asymptotic pdfs upon: (i) explicit reflection recipes for L\'{e}vy flights in bounded domains, \cite{dybiec}- \cite{zaba1}, (ii)  an impact  of  varied  reflection scenarios  in case of  the  fractional Brownian motions in a bounded domain, \cite{M,metzler,VHS}.

We stress that  the ultimate goal  of  computer-assisted procedures    is to get a reliable information about the asymptotic probability density, which is  inferred  path-wise, in terms of statistical data generated by the stochastic L\'{e}vy process,   in a suitable (time and space coarse-graining)  approximation.  That arises in  conjunction with   the stochastic differential equation (its random walk approximation), whose random variable respects    prescribed "reflection boundary"  properties.

 It is a priori not obvious,  whether or how  the path-wise  reflection scenario  induces  the Neumann-type boundary condition for the motion generator (e.g. the fractional Laplacian), \cite{gar,zaba}. In the present paper we favor the backward route and in selected cases, we   verify the validity of the presumed path-wise  reflection behavior of the jump-type process, whose    Neumann-nonlocally  constrained  dynamics of the probability density fucntions is    predefined, c.f.  Section V.

Our departure point is an observation that  the  physics-motivated research  is predominantly path-wise oriented, although   the   existence of the  stationary (steady  state) solution of the fractional Fokker-Planck equation  probability distribution   is  considered  as the major signature of confinement.   Shapes of corresponding pdfs, their peculiarities at the boundaries were analyzed both for  symmetric L\'{e}vy  processes and various (drifted) variants of the fractional Brownian motion. A common thread (rather operational input) in these research lines was a detailed path-wise  definition   of  the reflection  mechanism (procedure)  at a fixed boundary.

Somewhat interestingly, this viewpoint has not been shared by mathematically oriented scholars, and  no explicit  functional forms of    probability density functions,  {\it fully  consistent with}   (i) stochastic process with imposed  reflection conditions, (ii)  varied  domain restrictions for fractional Laplacians and  (iii)  Neumannn/reflection condition  proposals, can   found in the literature.

\section{Restricted versus regional fractional Laplacians: Whither the reflections are gone ?}

\subsection{Conundrum: What are stochastic processes connected with singular $\alpha $-harmonic functions ?}

The problem of steady-state Lévy flights in a confined domain has been addressed in Ref. \cite{denisov}  as that of the "distribution of symmetric Lévy flights in an infinitely deep potential well", the topic  which received  attention in connection with the  Dirichlet boundary data (admits killing  at the boundaries, but as well the inaccessible  ones, in reference to  taboo processes, \cite{gar}).  The term "infinite well" is somewhat   ambiguous and misleading.   There is an ample literature on the infinite potential well bound states for fractional Dirichlet Laplacians,  and the  related   issue of  spectral relaxation of  L\'{e}vy processes, c.f. \cite{gar,zaba,gar2} and compare e.g. \cite{dybiec,DNBD}.

The central result of Ref. \cite{denisov}, a specific probability distribution  in the interval $(0,b)$  (note that we refer to an open set, {\it not} to the closed one $[0,b]$)
  \be
P_{st}(x)=\frac{\Gamma(\alpha)(b)^{1-\alpha}[x(b-x)]^{\alpha/2-1}}{\Gamma^2(\alpha/2)},
\ee
 has  ultimately  (albeit not uncritically, \cite{gar,zaba})  received an   interpretation of the   statistical  signature of the  two-sided reflection of the  L\'{e}vy  process in the "infinite well". This interpretation is thought to be supported by  an analysis  (in part computer-assisted)  of  superharmonically  confined L\'{e}vy processes, \cite{kharcheva,dybiec,DNBD} and by properly engineered reflection scenario (stopping version of Ref. \cite{DNBD}, to be invoked  in below).

 Somewhat surprisingly, in Ref. \cite{denisov},   the validity of the exterior  Dirichlet condition has appeared    as the pre-requisite property for   would-be  "reflecting"  behavior. This in turn is  to be  enforced by impermeable boundaries.  The   Dirichlet regime surely stays  in line  with the previous wisdom gathered  for  infinite well spectral problems,  where the  exterior Dirichlet  restriction  has been directly  related to the L\'{e}vy  process with   killing and/or the problem of   barrier inaccessibility by the process, \cite{gar3}). The pertinent  spectral solutions  have been found    in  \cite{gar,zaba,gar2} (in part with computer assistance), and analytically in Refs.   \cite{dyda,kwasnicki,stos}.  Complementary discussions can be found in \cite{daoud,zaba1,kharcheva}. In reference to the interval problems, the pertinent eigenfunctions are  bounded and  continuous  up to the boundaries.  We note that these properties are not respected by the singular $\alpha $-harmonic function (9).

By  making a shift $x\rightarrow  x- b/2$  ($x=0$ is mapped into $-b/2$, while $x=b$ into $+b/2$)  and next  selecting  $b=2$, we can  replace   Eq. (9) by the form predominantly used in mathematical papers  \cite{dyda,Abatangelo1,daoud}, and likewise in \cite{gar,zaba1}.  The  closed  interval  of interest $[0,b]$   becomes $[-1,1]$, and its open version is $(-1,1)$.
In this notation, one can analytically demonstrate, \cite{dyda}  that  the fractional Laplacian, while acting upon  some functions, that are  identically  vanishing beyond  the open interval, produces  the value zero. (This property is related with the notion of  the  domain-restricted fractional Laplacian, c.f. \cite{gar})

Actually,  we deal with  a function $f(x)$,  defined on the whole of $R$,  which has the form  (up to a constant factor)
\be
f(x)=u(x)= (1-x^2)^{-1+ \alpha /2}
\ee
   if  $x \in (-1,1)$,  and identically  vanishes for all  $x\in R\setminus (-1,1)$. (We recall that the latter exterior condition has been employed in Ref. \cite{denisov}).

   We emphasize that  our function  is presumed to vanish  both  {\it  at}  the  boundary points (endpoints) $\pm 1$   and   {\it  beyond}  $[-1,1]$ as well.  This is the essence of the {\it exterior} Dirichlet boundary condition, which makes somewhat surprising the computational outcome,   confirmed analytically in Ref. \cite{dyda}  (see also section 5 of \cite{zaba1})
     \be
  (-\Delta )^{\alpha /2} u(x)= 0
  \ee
for  $u(x)$ of Eq. (10),  with  $x\in (-1,1)$.

An analogous to (11)  outcome is obtained for  odd   functions    $v(x)= x u(x)$, \cite{dyda}.  Functions that remain constant in $D=(-1,1)$ and vanish in $R\setminus D$, are valid elements of the (domain) kernel of the operator  $(-\Delta )^{\alpha /2}$ as well, compare e.g. also  \cite{gar}.

We mention that unbounded functions of the  form (9), (10)  have been  recognised  in the mathematical literature as  singular  $\alpha $-harmonic functions, and are  particular examples in a broader  family  of  "large"  and/or  "blow-up  solutions" of the fractional Laplacian equation, \cite{Abatangelo1}.  Interestingly, these functions were  introduced  without any association with the concept of reflecting L\'{e}vy processes in the interval, \cite{gar,Abatangelo1,ryznar}.

At this point we indicate the existence of a serious  drawback in both the analysis of \cite{denisov} and the ensuing  "reflective" interpretations  of singular $\alpha $-stable harmonic functions (9),(10).   The subject   of "stochastic processes connected with harmonic functions" has been addressed long time ago, \cite{elliott}, with the aim to classify   {\it various} examples of Cauchy processes constrained to stay in a compact interval $[0,a]$. This topic in more general $\alpha $-stable  context is still open,  specifically  as far as  the singular   $\alpha $-harmonic functions are to receive a consistent probabilistic interpretation.  Compare e.g. a pedestrian discussion in  sections 5 and 6 of Ref. \cite{zaba}.

We note in passing, that an asymptotic  accumulation of probability "mass"   at the domain boundaries, has been reported  for fractional Brownian motions with reflection, and  is known  to   produce  probability distribution  shapes closely related  to  these of  Eq. (9), \cite{M,metzler}. The accumulation effect  strongly depends on  a priori  prescribed  reflection scenario at the interval endpoints, and   - to the contrary -  we should keep in mind that  constant distributions  may be obtained  as well.

 Notwithstanding, these  observations  extend to L\'{e}vy processes with  two-sided reflections, see e.g. a computer-assisted analysis of Ref. \cite{dybiec,DNBD}, where the so-called "stopping" scenario  has been  activated in the Monte Carlo updating procedure,  at  the (small) distance $\epsilon $ from the boundary.

  This procedure has prohibited the L\'{e}vy process   from ever reaching or overshooting  the interval $[-1,1]$  endpoints, enforcing it to stay in the   $2\epsilon$-reduced {\it  closed}  interval  $[-1 + \epsilon ,1-\epsilon ]$ forever.   These $\epsilon $-reduced  boundaries  become  natural "stopping" points, where  "overshooting" jumps are interrupted, until the  jump away from the boundary is randomly sampled.   In Ref. \cite{DNBD},  the critical distance of the size  $\epsilon = 0.001$  has been employed, \cite{dybiec0}.

\subsection{Restricted fractional Laplacian.}

 Eq.(11), in view of the domain restriction (functions that vanish for all $x\in R\subset D$, $D=(-1,1)$),  refers to the so-called  {\it restricted}  fractional Laplacian, \cite{gar,kw,kw1}, for which we have coined the notational assignment $(-\Delta )^{\alpha /2}_{\cal{D}}$. Seemingly  this has   nothing   to do with Neumann boundary data and resultant (Neumann)  reflection scenarios.

For clarity of arguments, lets us recall, \cite{gar}, that  the  restricted   fractional Laplacian $(-\Delta )^{\alpha /2}_{\cal{D}}$,    shares an  integral definition   with     $(-\Delta )^{\alpha /2}$, c.f. Eq. (1),   but   normally   its  domain is supposed to   contain bounded functions only.  Thus,  the result (10), (11) goes beyond the standard framework, c.f. \cite{dyda,Abatangelo,Abatangelo1,gar}.

Anyway, for all $x\in D$  we have
\be
(-\Delta )^{\alpha /2}_{\cal{D}} f(x) = (-\Delta )^{\alpha /2}f(x)= h(x)
\ee
where   the function $h(x)$   may not share  the exterior     property  (vanishing  outside  $D$)  of  $f(x)$. 	We  point out  that there is no restriction   upon   the integration volume,   which is a priori $R$  and not  solely  $D\subset R$.

Remembering that  $p.v.$ indicates the Cauchy  principal  value of the involved integral, and that  $\mathcal{A}_{\alpha }$ has been defined in Eq. (8), we may write    for all $x\in D$:
\be
(-\Delta)_{\cal{D}}^{\alpha /2}f(x)=  \mathcal{A}_{\alpha }
\left[ p.v.   \int_{D}  \frac{f(x)-f(y)}{|x-y|^{\alpha +1}}\, dy  +   f(x)  \int_{R\backslash D} {\frac{dy}{|x-y|^{\alpha +1}}}  \right] .
 \ee
Here,  the exterior  $R^n\backslash D$  contribution  to the outcome of (12) has been  clearly  isolated.
 We point out that  the second term in Eq. (13) originally has contained a numerator of the form $f(x)- f(y)$, with $x \in D$  and $y \in R\setminus D$, which implies $f(y)=0$.

In passing, we  mention  another  minor conundrum,  which originates from well established properties of the  fractional Dirichlet Laplacian in a bounded domain $D$. Namely, this fractional operator   admits  a  solvable  spectral  (eigenvalue)  problem:
  $(-\Delta )^{\alpha /2}_{\cal{D}} \phi _k(x)    = \lambda _k \phi _k(x)$, with strictly positive  eigenvalues for all $k=1,2,...$.
  This   spectral  solution (with an   emphasis on  explicit eigenvalues and   eigenfunctions  shapes)  has  received  an ample coverage in the literature, c.f. \cite{gar2,kwasnicki} and  \cite{kulczycki}- \cite{zaba4}.  The  positivity of eigenvalues,  clearly stays at variance with Eq. (11), if spectrally interpreted.   Unless the notion of the singular $\alpha $-harmonic function is invoked, \cite{gar,zaba,Abatangelo1}.

\subsection{Censored L\'{e}vy process and the regional fractional Laplacian.}

     A censored stable process in an open set $D\subset  R$   is obtained from the
symmetric stable process by suppressing its jumps from  $D$  to the complement  $R \backslash D$
 of $D$, \cite{bogdan}. To this end one needs to restrict the  L\'{e}vy measure to $D$.  Told otherwise,
 a censored stable process in an open   domain D is a stable process forced  to stay inside D.   This makes a clear difference with a number of proposals to give meaning to Neumann-type  conditions, e.g. \cite{DRV,Abatangelo,Abatangelo1}, where outside jumps are in principle  admitted, albeit with an immediate return ("resurrection", c.f. \cite{bogdan}) to  the interior of $D$.

Verbally,  the censorship idea  resembles  that of   random  processes  conditioned to stay in a bounded domain forever, \cite{gar3,mazzolo}.  However,  the "censoring"  concept is not the same \cite{bogdan}  as that of the (Doob-type)   conditioning employed in \cite{gar3,mazzolo}.
   Instead, it is intimately   related to  reflected stable processes   in a bounded domain with killing within the domain, or in the least  at its boundary,  encompassing  a class of  processes  (loosely interpreted as "reflective")  that do  not approach   the boundary at all, \cite{bogdan,guan}.

In Ref. \cite{guan} the reflected stable processes in a bounded domain have been investigated,  stringent criterions for their admissibility set,   and their generators have been identified with  so-called {\it  regional} fractional Laplacians on the closed region $\bar{D}= D \cup \partial D$.
According to \cite{guan}, censored stable processes  of Ref. \cite{bogdan},  in $D$ and  for
$0<\alpha \leq 1$, are essentially the same as the reflected stable process.

In general, \cite{bogdan}, if $\alpha \leq 1$, the censored stable process is said to  never approach  $\partial D$. If $\alpha >1$, the censored process may have a finite lifetime and may take values at $\partial D$.

Conditions for the existence of the regional Laplacian for all $x\in \bar{D}$, need to be carefully set.
For $1\leq \alpha <2$, the existence of the regional Laplacian  for all $x\in \partial D$, is granted if and only if  a  derivative   (a  non-conventional   Neumann condition, that  is adapted to the nonocal setting) of a each function in the domain in the inward  direction vanishes, \cite{guan,guan1,ros}.

For our present purposes we assume $0<\alpha <2$ and  consider  an open set  $D\subset  R$.
The regional Laplacian is assumed  (a technical assumption employed in the mathematical literature) to act upon functions $f$  on an open set $D$ such that
\be
\int _D {\frac{|f(x)|}{(1+ |x|)^{1+\alpha }}}  \,  dx < \infty
\ee
For such functions $f$, $x\in D$ and $\epsilon >0$, we  write
\be
(-\Delta )^{\alpha /2}_{D,Reg}  f(x)  = \mathcal{A}_{\alpha }
\lim\limits_{\varepsilon\to 0^+}
\int\limits_{y\in D \{|y-x|>\varepsilon\}}
\frac{f(x)-f(y)}{|x-y|^{\alpha +1}}dy.
\ee
provided the limit (actually the Cauchy  principal value, $p.v.$) exists.

Note a serious conceptual and technical  difference between the restricted  and regional fractional Laplacians.
The former is restricted exclusively  by the domain property $f(x)=0$ for all $x\in R\backslash D$.  The latter
is   restricted    by demanding the  integration variable $y$  of the L\'{e}vy measure  to be in $D$, and  the domain restriction may or may not be introduced.

If we   actually   impose  the  exterior   Dirichlet  domain restriction     ($f(x)=0$ for $x\in R\backslash D$), then  Eq. (13) can be rewritten   as an  identity  relating the restricted and regional fractional Laplacians,  valid for all $x\in (-1,1)$,  \cite{bogdan,duo,gar}:
 \be
(- \Delta )^{\alpha /2}f(x)  = \left[ (-\Delta )^{\alpha /2}_{D,Reg}   +     \kappa _D(x) \right] f(x).
\ee
Here, for all $x\in (-1,1)$ we have   \cite{gar,duo}:
\be
\left[(- \Delta )^{\alpha /2} - (-\Delta )^{\alpha /2}_{D,Reg}\right] f(x)   =
  {\frac{\mathcal{A}_{\alpha}}{\alpha }} \left[ {\frac{1}{(1+x)^{\alpha }}} +{\frac{1}{(1-x)^{\alpha }}}\right]  f(x).
\ee
By invoking  Eqs. (10), (11) one may contemplate the differences    between  the restricted and regional fractional Laplacians, on a common (exterior) Dirichlet domain. Note the $\kappa _D(x)$ is positive and may be interpreted as a strongly confining perturbation (in $(-1,1)$) of the regional Laplacian. Consequently, the restricted Laplacian may possibly   be  interpreted    as the generator of a censored process with killing in $D$,  \cite{duo,bogdan}. The killing becomes strong in the vicinity of  boundaries, which stays at variance with any  probability  accumulation scenario therein.

In this (Dirichlet) regime, we readily see that  the singular harmonic function  (10) is not a solution  of  $(-\Delta )^{\alpha /2}_{D,Reg} f(x) = 0$  for $x\in  D =(-1,1)$.    Hence, if (10) is to be interpreted as the outcome of  the $\alpha $-stable  L\'{e}vy process with   two-sided reflections, the regional Laplacian is surely not the generator of such  random  process. The reasoning of Refs. \cite{bogdan,guan,guan1} does not encompass this case.

\section{How can one  "see"  L\'{e}vy random variables and    processes in  the  "reflecting"  interval  ?}
\noindent

\subsection{Visualisation method in $R$.}
Let  $\{X(t),t\geqslant 0\}$ be any    $\alpha$-stable Lévy  process with $\alpha \in  (0,2]$.
We stay within the ramifications of   Section 1 and consider symmetric $\alpha $-stable processes on $R$, with  probability density functions   encoded in the notation  $X\sim S_\alpha(1)$.  The general  trajectory (sample path) generating algorithm,  originally formulated for    general  $\alpha $-stable  processes, and codified in Ref. \cite{bratley,JW}, see also \cite{JW1,JW2,AG}, for symmetric processes   takes a considerably  simpler form.

The algorithm is composed of two   steps.   First we generate  a random variable $V$
from the uniform probability distribution on  $\left(-\frac{\pi}{2},\frac{\pi}{2}\right)$, together with a random variable  $W$ form the exponential distribution with the mean value $1$.  The next step of the algorithm amounts to
the evaluation of
\be
X = \frac{\sin(\alpha V)}{(\cos V)^{1/\alpha}}\cdot\left\{\frac{\cos(V-\alpha V)}{W}\right\}^{(1-\alpha)/\alpha}.
\ee
So defined random variable $X$ is the $\alpha $-stable one and $X\sim S_{\alpha }(1)$, \cite{JW}.

At this point we recall  the  scaling   property  $X\sim S_\alpha(1)$ implies $Y=\sigma X \sim   S_\alpha(\sigma )$.  Since the process has independent increments, we have a clear path towards defining the displacements in time (or whatever parameter that might play this role), and thus a simulation of   L\'{e}vy random walk:
\be
X(t)-X(s)\sim S_\alpha((t-s)^{1/\alpha}),
\ee
for all  $0\leqslant s<t<\infty$.

 The $s$, $t$ labels need not to be continuous. In particular,
  setting $t=N$, and presuming that an  initial value of $X(0)$  is a priori chosen,  we can re-interpret $X(t)$ with $t\in [0,T]$  as  an  exemplary   L\'{e}vy walk started at $X(0)$,  and  terminated  at $X(T=N)$  after $N$ random jumps. Accordingly,
   \be
  X(t) \longrightarrow  X(N)= \sum_{k=1}^N  Y(k) + X(0).
  \ee
 where $N$  consecutive (random) displacements   $Y(k)=X(k)-X(k-1)$  are  sampled according to   $Y(k) \sim S_{\alpha }(1)$  for all $1\leq k\leq N$.

We note that, if  to  insist on  $t$  {\it  not  to be } a natural number,   the process   $X(t)$ may still  be represented as a sum of random variables  $\sim  S_\alpha(1)$ plus  another  random  variable  $\sim  S(\gamma^{1/\alpha})$, where  $0<\gamma=t-\lfloor t \rfloor<1$ ($\lfloor t \rfloor = \max\{k\in\mathbb{Z}: k\leqslant t\}$).  See e.g. the property (19).

Let us describe  how to  handle an arbitrary time    interval   $t \in [0,T]$ in the construction of  the  L\'{e}vy random walk. We divide $[0,T]$ into $2^N$ equal pieces. For each $i$-th  segment $[t_{i-1},t_i]$, where $i=1,2,... 2^N-1$ and $t_0=0$, $t_{2^N}=T$, we generate  the random variable $X \sim S_{\alpha }(1)$, by means of   Eq. (18). Since $X\sim S_{\alpha }(1)$  implies $Y= \sigma X  \sim S_{\alpha }(\sigma)$, we realize   that    the  random variable
\be
Y(i)=  (t_{i} - t_{i-1})^{1/\alpha } \, X \sim   S_{\alpha }\left({\frac{T}{2^N}}\right)^{1/\alpha }
\ee
shares a probability distribution       with  $X(t_i)- X(t_{i-1})$, compare e.g. Eq. (19).
This  distribution is common for all random increments  $Y(i)$, indexed by $1\leq i \leq 2^N$.

\subsection{The  Skorohod reflection scenario.}

We have mentioned before that  the  very  concept of reflection from the barrier is  still under dispute in the literature on   $\alpha $-stable  L\'{e}vy  processes. Nonetheless, there is a classic proposal, introduced by Skorohod, \cite{skorohod,pilipenko} which (to our surprise) seems to have never been explicitly used in the physics-oriented research, \cite{dybiec,DNBD,denisov,M,VHS},  and  has been  rather seldom mentioned in the math-oriented papers  (a notable exception is the series of publications by Asmussen and collaborators, \cite{AG,As,As1,As2}. see also \cite{KLRS,ibrahimov,ievlev}).

Let $\{X(t),t\geqslant 0\}$ be  a symmetric  $\alpha$-stable L\'{e}vy process on R.  We want to deduce its version, which is confined in the closed interval $[0,b]\subset R$  by a two-sided reflection from endpoints. To give meaning to the term "reflection", we shall rely on the Skorohod proposal \cite{skorohod} on how to implement a reflection from a single barrier, which can be readily  extended   to  the two-barriers (endpoints of $[0,b]$) case.  We denote  $R(0) =X(0)$  the a priori chosen  initial value, which we associate with the initial time instant $t_0=0$.

We denote  $\{ R(t),t\geqslant 0\}$  the   jump-type process that is entirely contained in $[0,b]$,   and  formally  defined as follows,  \cite{As}:
\be
R(t)=R(0)+X(t)+L(t)-U(t),
\ee
where  $L$ and  $U$  are non-decreasing, right continuous   {\it compensating} jump-type processes such that
\be
\int_0^\infty R(t)dL(t)=0,\quad \int_0^\infty (b-R(t)) dU(t)=0.
\ee
Given  the $\alpha $-stable process  $X(t)$, we say a triplet $\{ R(t), L(t), U(t)\}$ of processes is a solution of the {\it Skorohod problem}  on $[0,b]$, if $R(t) \in [0,b]$ for all $t$. The mapping, which associates  the above triplet to $X(t)$ is called the Skorohod map, \cite{As}.

The integral conditons upon {\it regulators}  $L(t)$ and $U(t)$,  \cite{As} (we prefer to name them compensating processes), secure that $L$ can only increase when $R$  actually  is at the lower boundary $0$, and $U$ only
when $V$ is at the upper boundary $b$.    Thus,  loosely speaking,  $L$ represents the "pushing up from $0$" that
is needed to keep $R(t) \geq 0$ for all $t$, and $U$ represents the "pushing down from $b$"
that is needed to keep $R(t) \geq b$ for all $t$.
 Since $X(t)$  proper   has a  jumping distribution with unrestricted jump sizes, $L$ becomes activated when $X(t)$ would overshoot the barrier $0$, and $R(t)$ would have to pass $0$ to the left (becoming negative). Analogously $U$ becomes activated , if the $X(t)$ overshooting would make $R(t)$ to pass $b$ to the right.

The compensating role of $L$ and $U$ may be easier to decipher,   while passing to the random walk approximation of the Skorohod reflecting process. This allows to  bypass  intricacies related to the continuous case, \cite{As}-\cite{ievlev}.

The random walk with a two-sided reflection  in $[0,b]$ we define as follows, \cite{As}:
\be
R_n = \min\left(b,\max\left(0, R_{n-1}+ Y_n\right)\right),
\ee
where random variables  $Y_n$,  see  e.g. (20) and  (21), are  inferred from $X(t)$ and have identical probability distributions.  We set the starting point of the walk $R_0=v$, where   $v\in[0,b]$.

The original L\'{e}vy process is thus replaced  by the random walk   $X(n)$ of Eq. (20)   where $Y(i)$'s have identical distributions, c.f. Eqs. (19)-(21).  For computational convenience, we shall  choose  $Y_n=X(n)-X(n-1)$  to have a distribution  $S_\alpha(1)$  for all $n$.  This suffices to achieve  a qualitative picture of large $n$ (asymptotic) probability distributions of the random walk, where merely a large number of (time) steps matters, and not their specific size (duration).

  We point out that  to improve  an approximation finesse  of $X(t)$ by the random walk,  one should assume that  $Y_n\sim S(\sigma)$  with  $\sigma<1 $ (eventually  $\sigma \ll 1$).   This can be always accomplished by employing  the  scaling   property  $X\sim S_\alpha(1)$ implies $Y=\sigma X \sim   S_\alpha(\sigma )$.

In below we shall test the case $Y_n\sim S(1)$,  to generate  long (walk "time")  sample  trajectories on  $[0,b]$, actually  with $T=2500$.  Before passing to inferred probability distributions (rather   histograms obtained from statistical data   for $500 000$ sample paths of the considered reflecting random walk), we shall analyze in more detail the role of $L$ and $U$ compensating processes (regulators), that transform the unrestricted  $\alpha $-stable  random walk into a  reflecting one  within  $[0,b]$.

\begin{figure}[htp]
\begin{center}
\centering
\includegraphics[width=90mm,height=80mm]{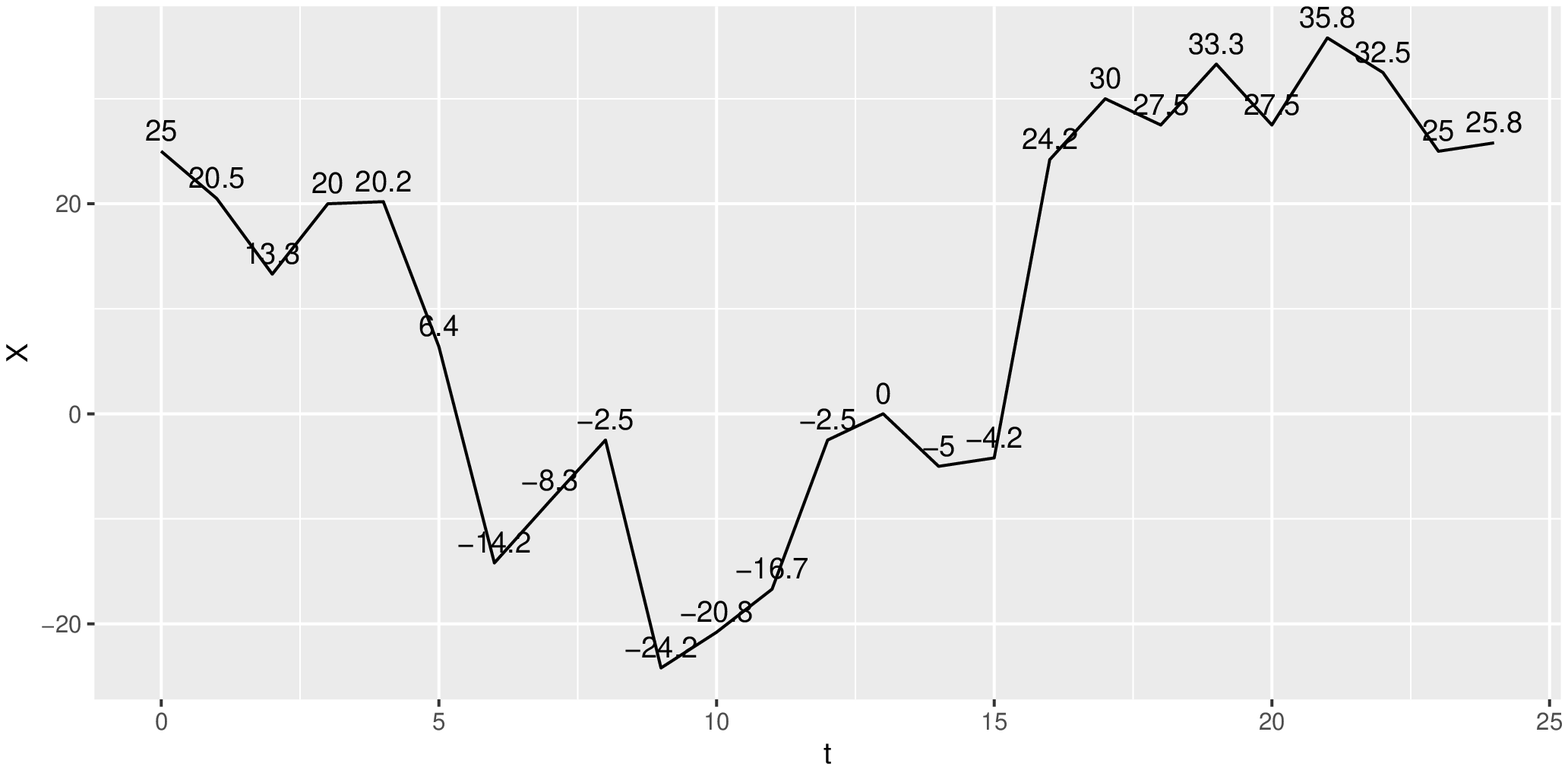}
\includegraphics[width=90mm,height=80mm]{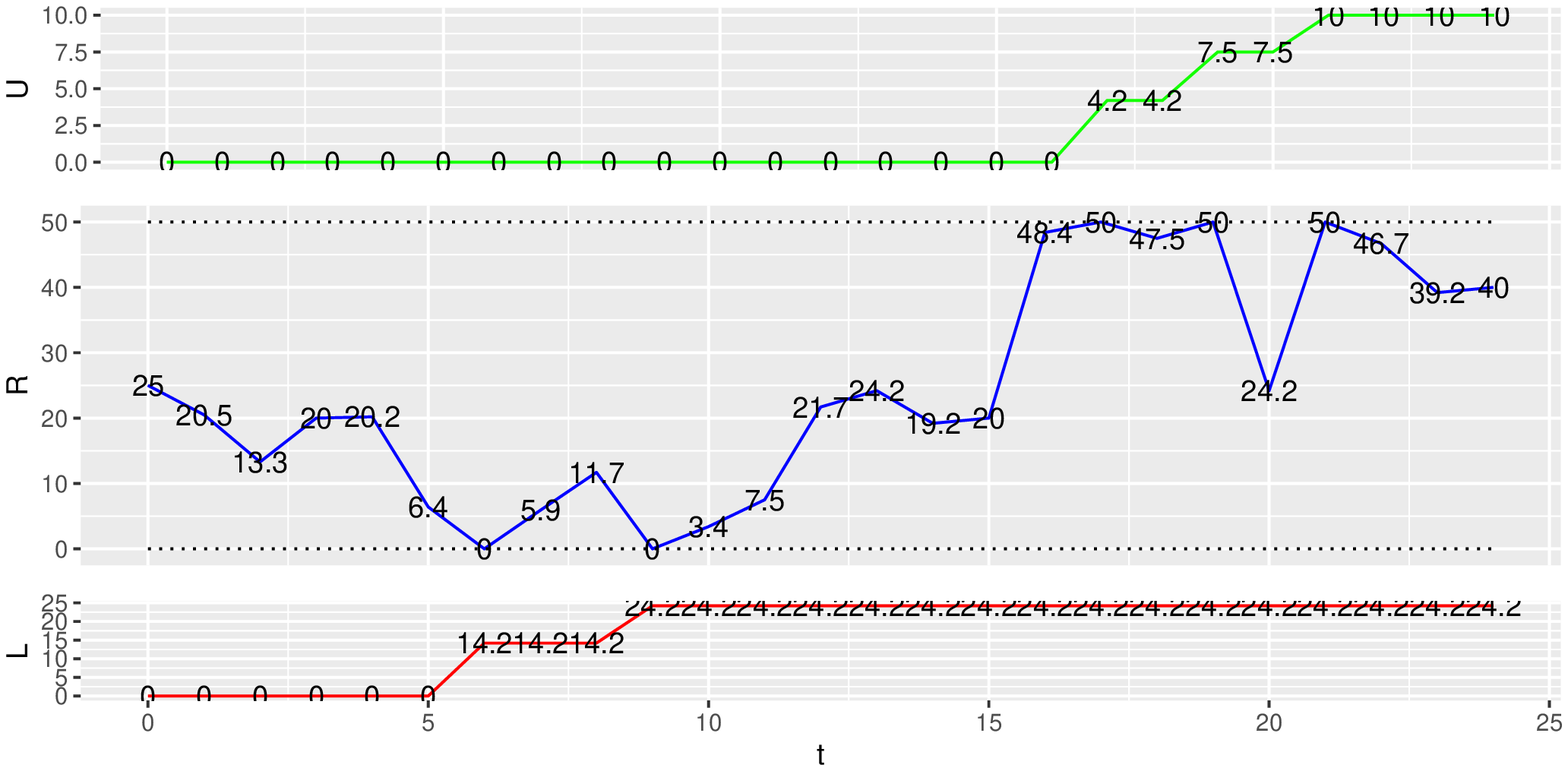}
\caption{The Skorohod map in terms of random walks. Left panel: unrestricted L\'{e}vy walk sample trajectory with $R(0)=X(0)=25$ and the number $N=24$ of   consecutive  "time" steps (e.g. displacements $Y(i)$).  Right panel depicts a construction of the Skorohod triplet  $R, L, U$, where $R$ stands for a reflecting L\'{e}vy random walk in $[0,50]$, started at $X(0)=25$.  The  compensating walks (e.g. lower $L$ and upper $U$ regulators) are inferred from the  data of the unrestricted process, by means of  the  recursion (24).}
\end{center}
\end{figure}

The Skorohod map  is detailed  in Fig. 1. The reflecting L\'{e}vy  walk  is inferred from the  standard unrestricted L\'{e}vy walk, as introduced in Eq. (20), in accordance with the concept of the Skorohod map.
  We note that integer labels $k$ effectively correspond to  normalised (length $1$)  "time" steps. Indeed, it is enough to consider a test  time interval $[0,T]$, which is composed of   $N$   integer "time" points, so that  we deal with  normalized  increments $t_{i}-t_{i-1}=1$ for each value of  $1\leq i \leq N$. We note that the scaling (19) is  here immaterial, because $(t_i-t_{i-1})^{1/\alpha } =1$.

We choose $N=24$ and create a sample trajectory of the {\it unrestricted} L\'{e}vy walk starting from the value  $R(0)=25$, c.f. (24). Since $X(1)=20.5$, we get $Y(1)=-4.5$. In the left panel of Fig.1 the sample path of our walk is represented by randomly assigned   jumps, which produce coordinate labels with values along the  vertical axis.  So e.g. consecutive jumps $-4.5;-7.2;+6.7;+0.2;-13.8;$ etc., give rise to the  {\it unrestricted}   random walk steps  $20.5;13.3;20;20.02;6.4$ etc., as depicted in the  panel.

To realize the Skorohod map, we actually need to properly "tailor" the obtained unrestricted sample trajectory, so that it will fit to the a priori chosen residence interval $[0,b]$ with $b=50$ of the prospective  two-sided  reflected L\'{e}vy walk. This is detailed in the right panel of Fig.1, in terms of two compensating random walks: (i) the    upper barrier regulator  $U(N)$ (green) and  (ii) the lower barrier regulator   $L(N)$ (red).

To implement the reflecting walk we employ the recursion formula (24), where the regulators are visually absent. However, we can reconstruct  them (this is the essence of the Skorohod map) as follows.

 In the $6$th  displacement step,  the unrestricted process induces a jump $-20.6$    from $6.4$ to $-14.2$. According to Eq. (24) the jump is interrupted (stopped) at $0$,  while the remaining part $-14.2$ must be compensated by the increase (jump  from $L(5)=0$ to $L(6)=+14.2$ of the regulator $L$, which is depicted in red in the right panel. We have reached the value $R(6)=0$, beginning  from which  two consecutive  jumps of the unrestricted process,  $+5.8$ and $+5.9$,  do not induce any increment of $L$, and  ultimately $R(8)= 11.7$.

  However the trajectory of the unrestricted process,  at the $9$th  "time" step jumps down by $-21.7$. On the level of  reflected process $R$, according to Eq. (24),  such jump  from the value $R(8)=11.7$   is interrupted (stopped)  at the value  $R(9)=0$,  and  the regulator $L$ is activated again. This means that the  compensating jump of the size  $21.7 -11.7=10$ needs to be executed, thus  setting the value   of the lower regulator at  $L(9)=24.2$. This value is preserved up to $L(24)$.

  We proceed analogously with the upper regulator, whose value equals zero up to the $17$th "time" step, when the "overshoot" occurs and  upper value $R(17)= 50$ is reached while executing the jump $+5.8$ from $R(16)=48.4$.
  The overshooting surplus $5.8- 1.6=4.2$  is depicted as the value $U(17)$. Further procedure follows analogously.

\begin{figure}[htp]
\begin{center}
\centering
\includegraphics[width=80mm,height=80mm]{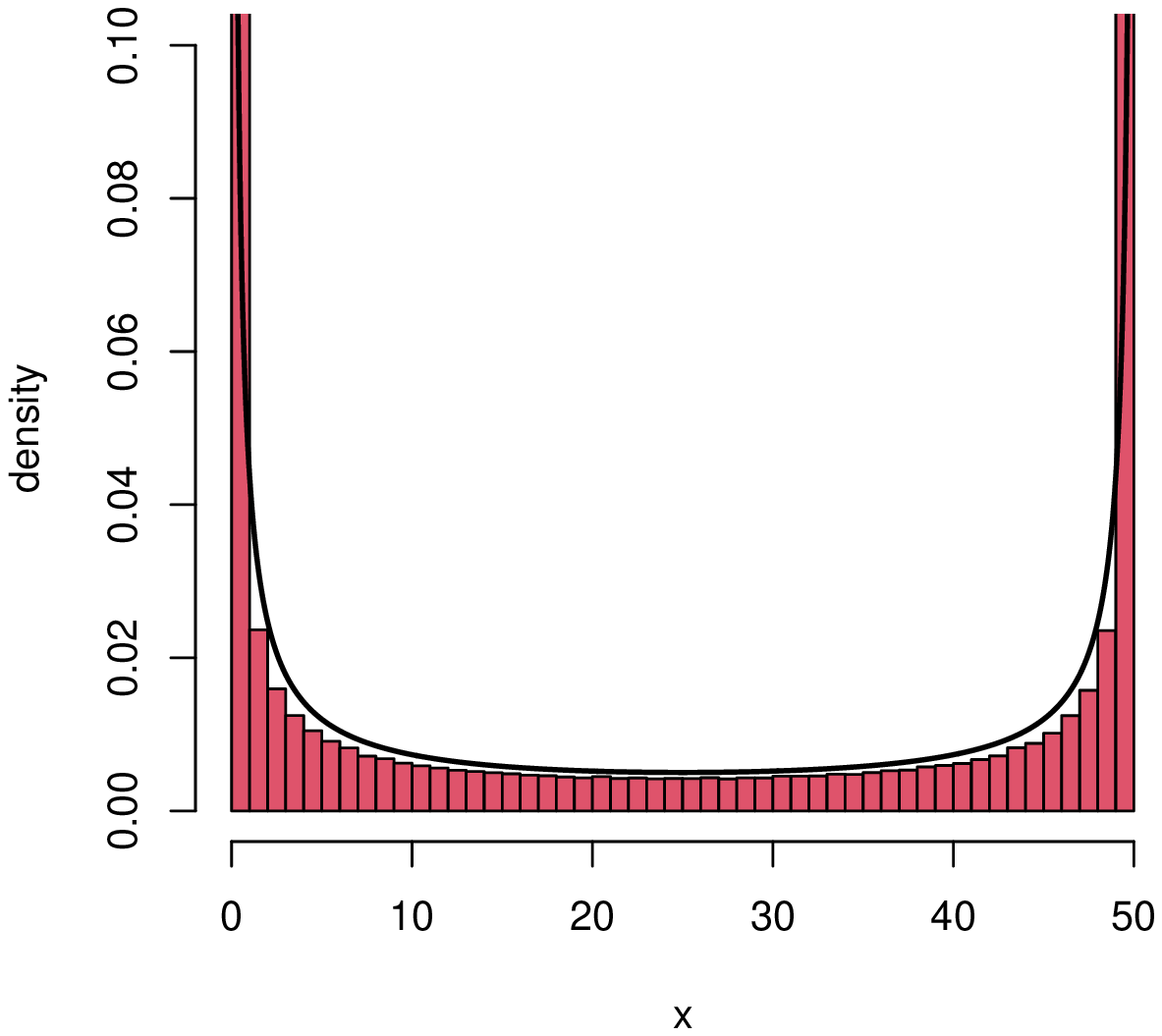}
\includegraphics[width=80mm,height=80mm]{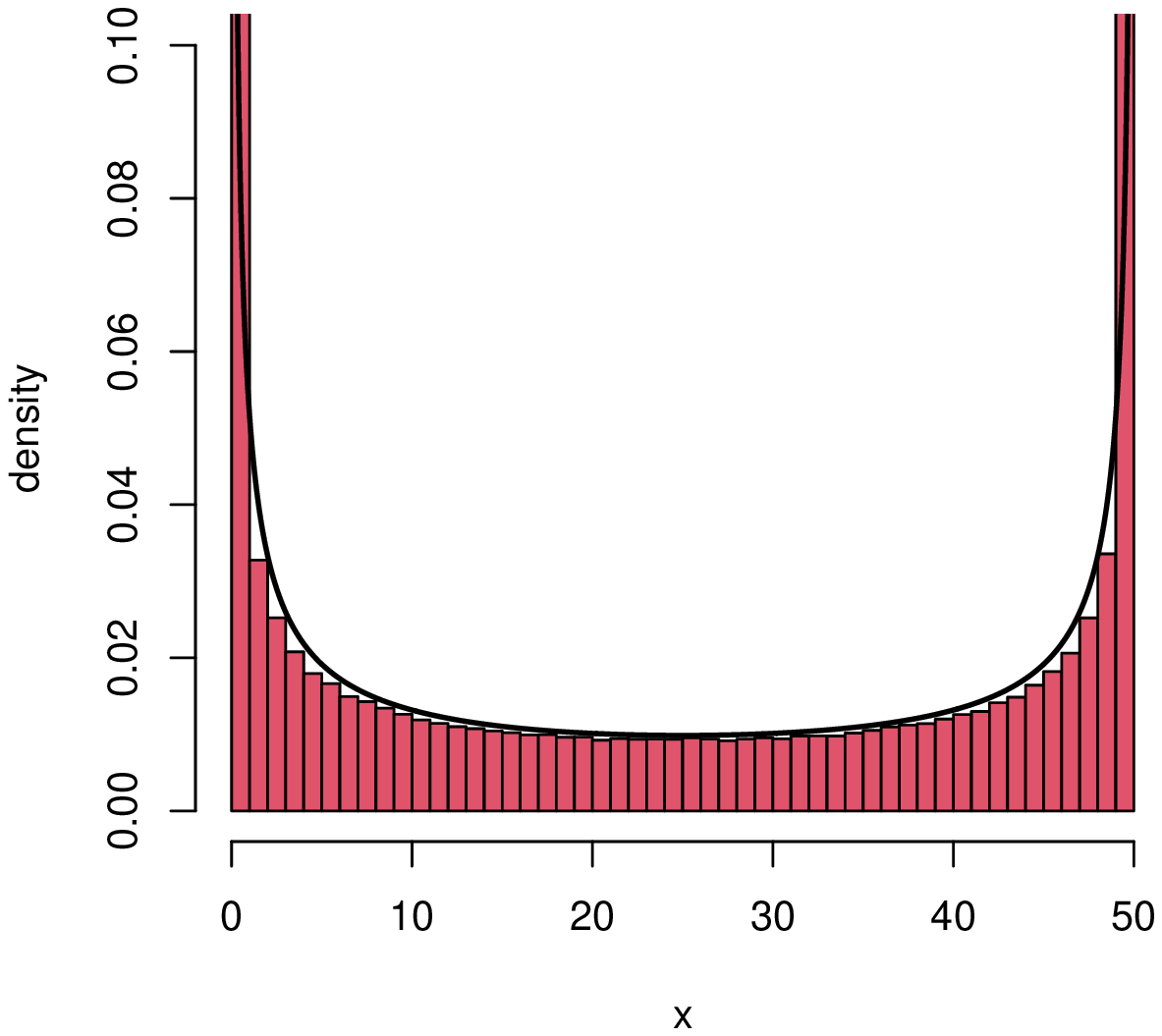}
\caption{Skorohod random walk. Statistics (histograms) of $500 000$ sample trajectory location "hits" at  $T=2 500$, for  $\alpha=0.3$ (left panel)and $0.7$  (right panel). All trajectories were started at $X(0)=25$.   A continuous best fit  curve, actually  coincides with the singular $\alpha $-harmonic function  of  Eqs. (9), known to solve Eq. (11)  in the (open) interval $(0, b=50)$, c.f. \cite{dyda}.}
\end{center}
\end{figure}

\subsection{Skorohod random walk. Large time asymptotics and approximate stationary probability densities.}

We are interested   in statistical properties (strictly speaking, in inferred probability density  functions), induced by the discretized (random walk) representation of solutions to stochastic differential equations with two-sided reflection constraints. To this end we divide the time interval $t\in [0,T]$ into $N$ equal segments, so that each time instant  $t_i>0$ is a natural number, and  $t_{i}-t_{i-1}=1$ for all $i$. This allows to neglect scaling issues of Eq. (19).
We are interested in the (sufficiently) large time asymptotic, hence in  statistical data (histograms of trajectory hits) for  large values of $T$. Our test choice has been $T=2 500$.   We remember that  for  each spatial  increment, see e.g. Eqs. (20), (21), there holds  $Y_i \sim S_\alpha(1)$.

 In Figs. 2-4  we depict the statistical data (histograms) at $T= 2500$, inferred  for  $500 000$  sample  trajectories  of the Skorohod random walk  in the interval $[0,50]$,  all started at $X(0)=25$,  and  generated for selected values  $\alpha=0.3, 0.7, 1, 1.3, 1.7$  of the  stability parameter. The continuous curve depicted in black,   appears to be a  definite best fit  probability density function, actually coinciding with the singular $\alpha $-harmonic function of Eqs. (9), c.f. also  (11).

A compatibility of  histograms for the Skorohod random walk in the  reflecting interval  with the approximating continuous curve (9) is quite satisfactory, even for normalized time increments. The approximation finesse can be easily improved, if to generate trajectories with time increments significantly less than one  (we need $\sim  S_\alpha(\sigma)$, $\sigma <1$). This would increase the computer time cost, even  while keeping  fixed  $T=2500$.

  We have  explicitly  tested the finesse improvement, by  passing to  the time increment equal $1/8$. The comparative outcome is depicted in Fig. 3  for  $\alpha=1.7$ and  $T=2500$. An approximation accuracy improvement can be visually verified.  Analogous  comparative  tests   are depicted  in Fig.4  for $\alpha = 0.3, 1, 1.7$.

 \begin{figure}[htp]
\begin{center}
\includegraphics[width=80mm,height=80mm]{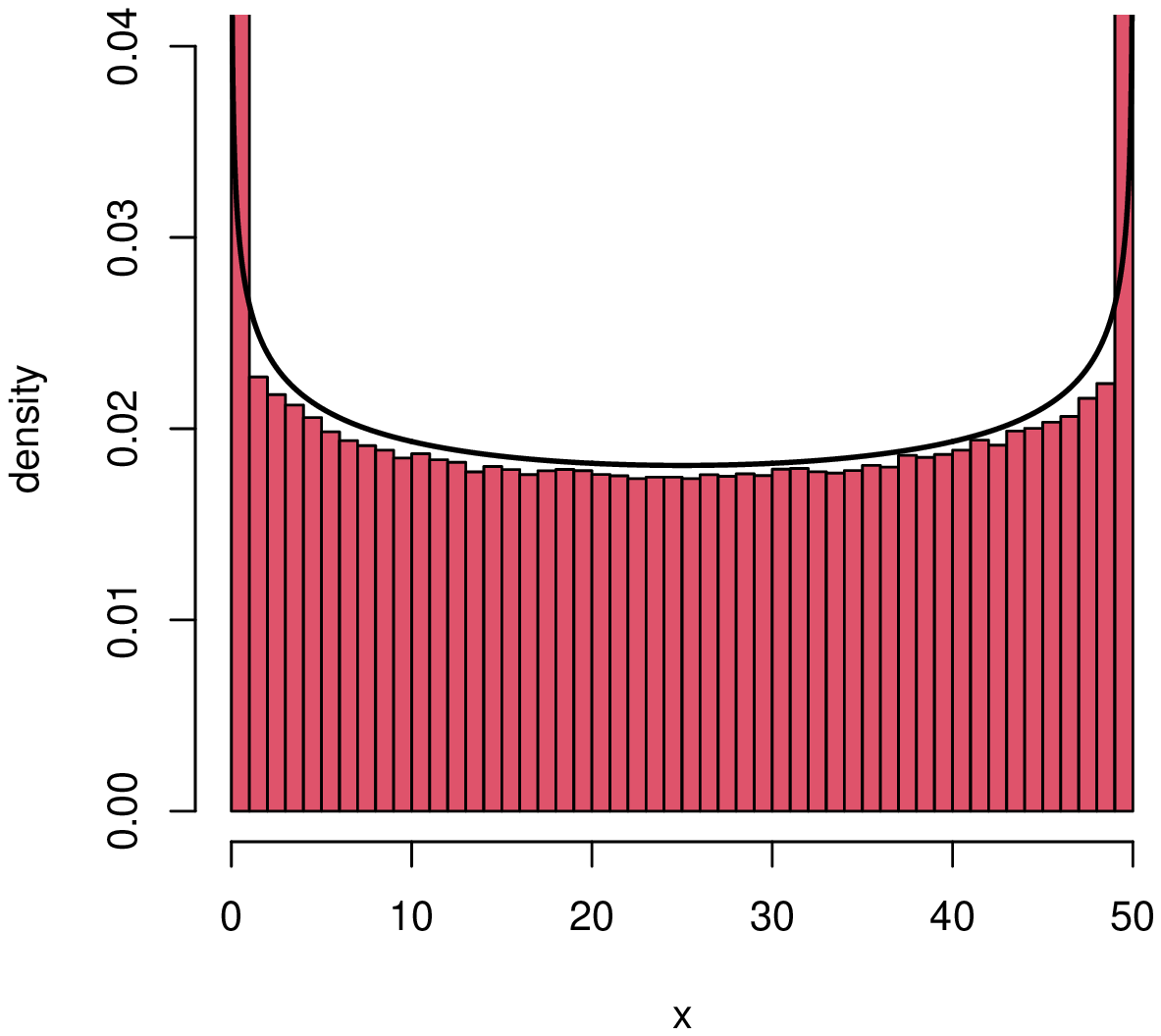}
\includegraphics[width=80mm,height=80mm]{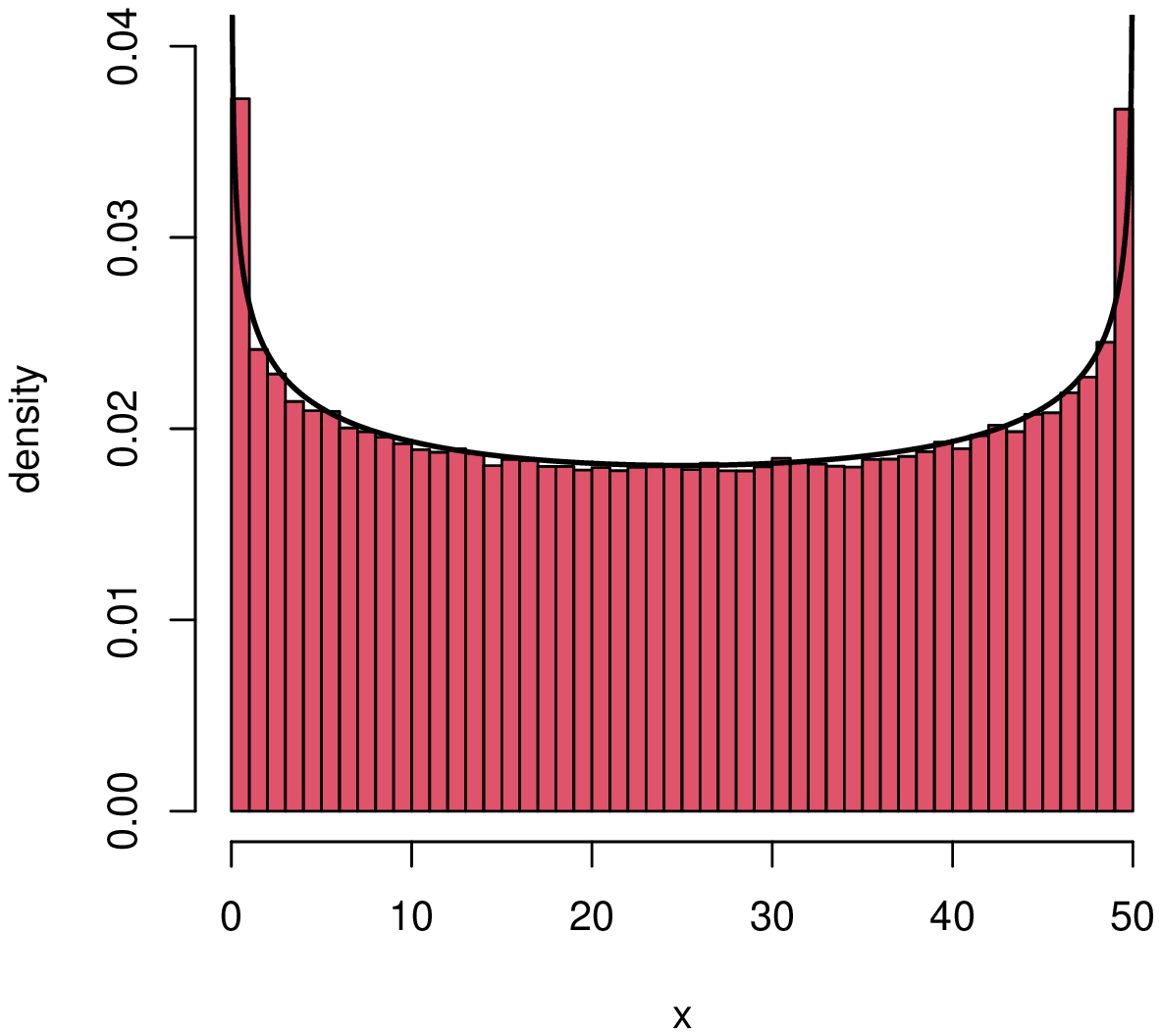}
\caption{Statistics (histograms) of  $500 000$  trajectories generated for  $\alpha=  1.7$, using a normalised (equal $1$ in the left panel)  time increment, and comparatively (to indicate a visually significant increase of the approximation finesse)  for its $1/8$ version  (right panel).  A continuous curve coincides with this of Eq. (9), for $b=50$.}
\end{center}
\end{figure}

\begin{figure}[htp]
\begin{center}
\centering
\includegraphics[width=80mm,height=80mm]{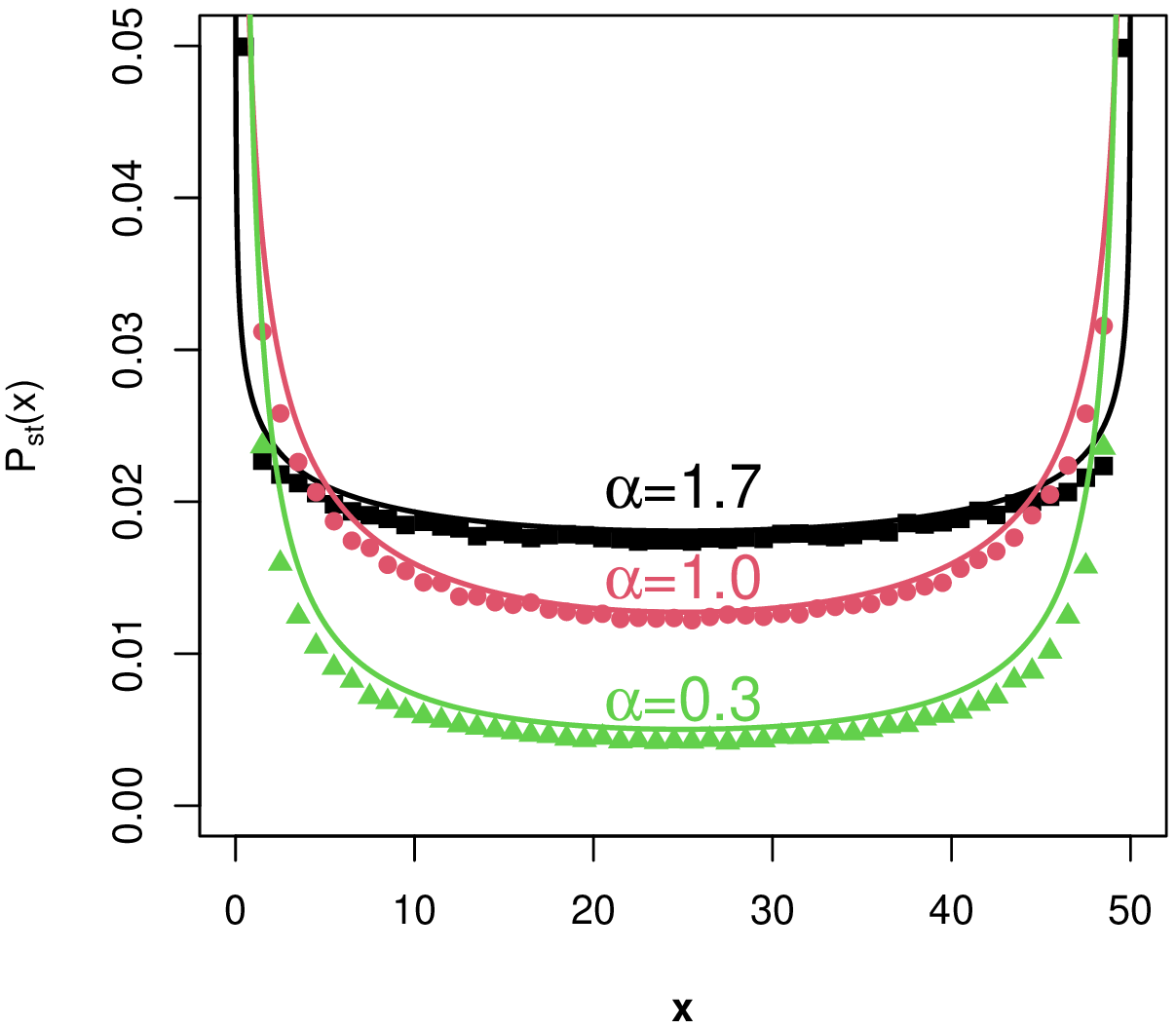}
\includegraphics[width=80mm,height=80mm]{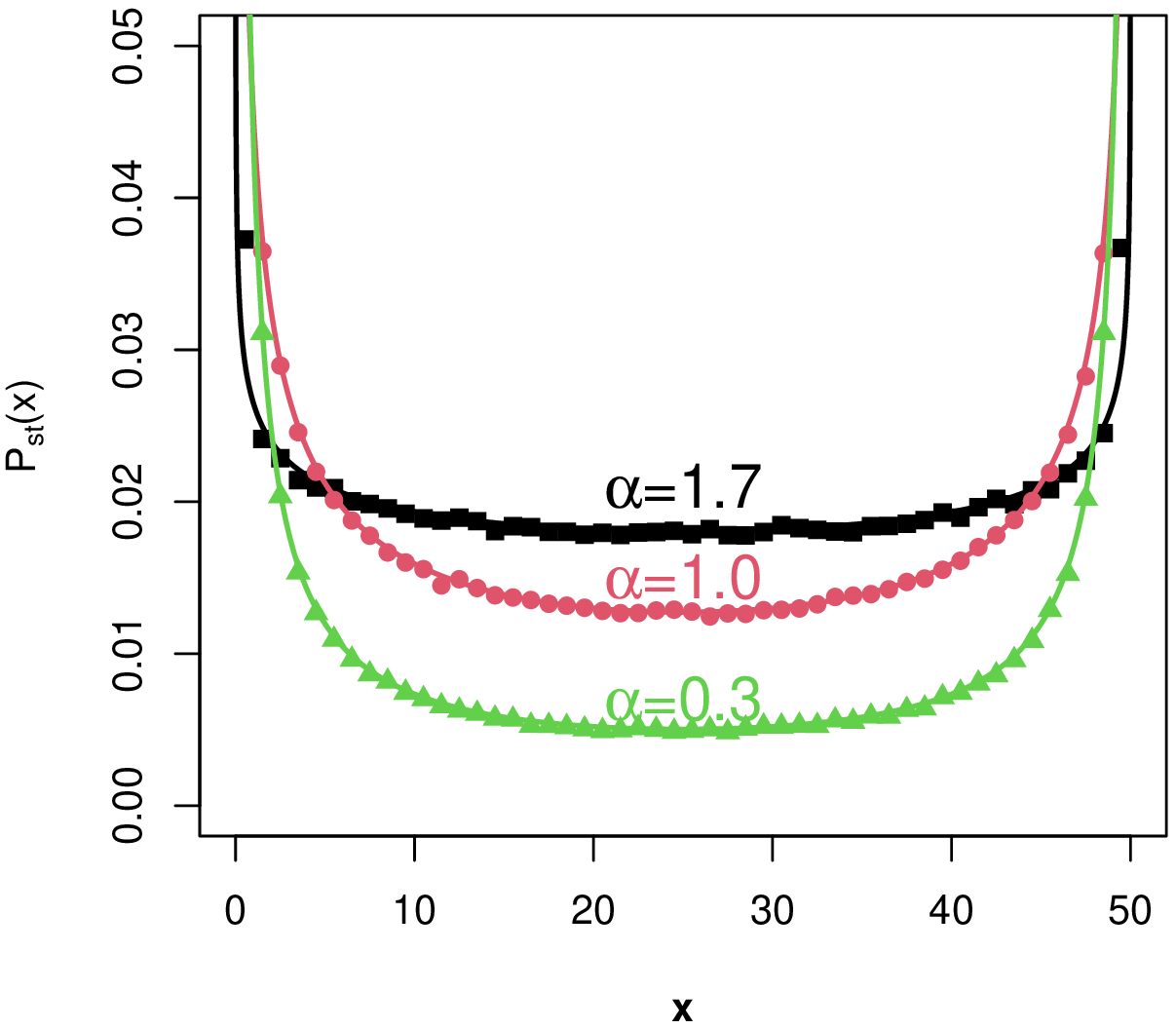}
\caption{Skorohod random walk.  A comparative statistics for  $500 000$ trajectories, generated for   $\alpha=0.3, 1, 1.7$, with the unit time increment (left panel) and its $1/8$ version. Coordinates  of each displayed dot (triangle, square, circle) are specified as follows:  abscissa axis specifies histogram midpoints, while the ordinate axis  refers to  probability density (actually the frequency of trajectory  "hits"). The starting point $X(0)=25$ is common for all trajectories.  The continuous  curve (9) is displayed  in black.}
\end{center}
\end{figure}

\section{Beyond the  Skorohod map. Alternative   reflection  scenarios at a  barrier.}

\subsection{Interception, or stopping at the barrier.}

The discretisation of the form  (24) of $R(t)$, without explicit  reference to the Skorohod problem, can be found as  Eq. (10) in Ref. \cite{VHS},  in  connection with the  reflection scenario  from a single barrier located at $w>0$, with the  forbidden region $x<w$  (actually  $w$  can be identified with $0$).  Then   the recursion,  originally    carried out   with reference to the  increments $\xi $ of the  fractional Brownian motion, \cite{VHS},   (with no abuse of notation we can interpret $\xi $ as increments of the  standard  Brownian motion or any $\alpha $-stable L\'{e}vy process)

\be
x_{n+1} = \max (x_n + \xi_n, w)
\ee
places the particle right at the barrier,  if the step would take it into the forbidden region $x < w$.    Clearly, we encounter an interception of the "overshooting" jump by the barrier. This (barrier) stopping location is the starting point for the consecutive  jump. If  the jump  would  potentially  take the trajectory to the forbidden region, it is not realised and the barrier location remains the stopping point, until a "proper" jump  gets sampled.

 The definition  (24) has been interpreted as a discretised version of the reflecting behavior in the  fractional Brownian motion (FBM). It is often employed in the mathematical literature, in the context of  queueing theory \cite{FBM,FBM1}, and  likewise   for  L\'{e}vy  processes, \cite{As,As1,As2}. \\

Reflections  (25) from the bottom barrier $w \in R$,   can be readily transcribed to the L\'{e}vy setting by means of obvious substitutions  (compare e.g. Eq. (24)), resulting in:
\be
R_n= \max (w,  R_{n-1} + Y_n)
\ee
 Formulas for the upper barrier readily follow.  Eq. (24) actually combines the bottom and upper barrier cases in a single  recursion  formula, for the  L\'{e}vy  walk with reflections at endpoints of the interval,  $[0,b]$, where $w=0$ stands for the lower barrier.  This is consistent with the  two-sided  (Skorohod)   reflection scenario  and the asymptotic behavior of probability distribution inferred from the statistics of  a large number of random  paths, as reported in Figs.  2, 3 and   4, with an approximating pdf  of Eq. (9).\\

{\bf Remark:}
 Under the name of the "motion  stopping"  scenario, a very similar proposal has been employed in Ref. \cite{dybiec}.
 In  reference to the  interval $[0,b]$, a trajectory that crosses $0$ is paused at   $+ \epsilon >0$, which  is supposed to be  small. The point $+\epsilon $ is used as a starting point for the next jump.    Clearly, if the  consecutive  jump would possibly "overshoot"  $+\epsilon $ in the negative direction, the motion  would remain  stopped until the   move in the positive direction is enabled again.
  A closer examination  proves that this   $\epsilon $-stopping  scenario is equivalent to the Skorohod random walk in the $2\epsilon $ - reduced interval $[\epsilon, b- \epsilon ]$. In Ref.  \cite{dybiec}, the Authors have chosen $\epsilon = 0.001$, \cite{dybiec0}. Monte carlo simulations have confirmed that the curve (9) is a reliable  continuous  approximation of the statistical (histograms) data for asymptotic probability distributions in the pertinent random walk.

\subsection{Wrapping and mirror reflection.}

 Another reflection scenario proposal, restricting the random walk to $x \geq w$ can be defined by means of a recursion:
\be
x_{n+1} = w + |x_n + \xi _n -w|,
\ee
which for $w=0$ is recognizable as the  standard (Brownian by origin) reflection  from an "elastic" wall. For $w>0$, the reflection formula (27) describes the {\it wrapping } scenario , c.f. \cite{DNBD}, where a  sample  trajectory that  would potentially end at $x<w$, actually  is wrapped around $w$ and the "jump length surplus" $|x-w|$ is added to $w$ to get the final outcome (e.g. reflection through wrapping). We note that for $w=0$,  and $x_n >0$, we have $x_{n+1} = |x_n + \xi _n|$, which is a mirror reflection at $0$.  \\

{\bf Remark:} We recognize in Eq. (26) the wrapping scenario of reflection from the lower barrier,  originally adopted to the interval $[-L,L]$, in which the    $\alpha $ -  stable process was supposed to be confined. In this case, the numerically assisted statistical analysis, has revealed that the asymptotic probability density function needs to be a constant, \cite{dybiec0}.\\

For the L\'{e}vy walk subject to the mirror reflection scenario at the endpoints of $[0,b]$, we   shall  explicitly  demonstrate signatures of  convergence    to a constant  distribution in  Fig. 4. That remains  in conformity   with the spectral analysis of the regional fractional Laplacian, outlined in Ref.  \cite{gar}, see also \cite{ievlev}.

\begin{figure}[htp]
\begin{center}
\centering
\includegraphics[width=70mm,height=70mm]{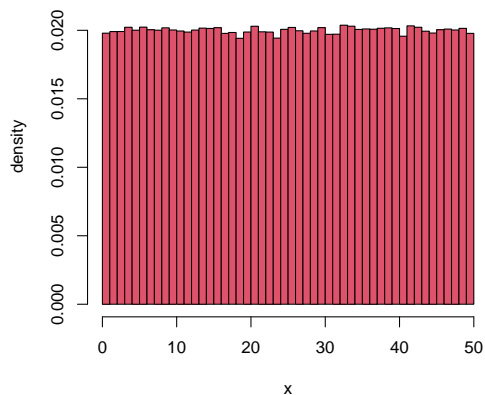}
\caption{Mirror reflection  in $[0,50]$. Statistics of  $500 000$ trajectories  for  $\alpha=0.3$ and the normalized time increment. All trajectories have   a common starting point  $X(0)=25$.}
\end{center}
\end{figure}

For the reader convenience, we outline our procedure in the mirror case, when the lower barrier is set at $0$,  the interval of interest is $[0,b]$, with $b=50$, and  $X(0)=25$ is  chosen as  the starting point of the random walk.

For the $\alpha $-stable L\'{e}vy process, we define the  mirror reflection from the lower barrier at $0$ as follows:
\be
R_{n}=|R_{n-1} + Y_n|.
\ee
Here   $n=1, 2, \ldots$, and   $R(0)=X(0)= b/2$, compare e.g. \cite{VHS,DNBD}.
The upper barrier we set at $b$.  To  arrive at  a mirror reflection,  we   modify appropriately the reflection  recipe (27):
\be
R_n  = w - |R_{n-1} + Y_{n} - w|
\ee

To arrive at a consistent    two-sided reflection in  $[0,b]$, we need to keep  resultant $R_n \in [0,50]$, while remembering that   the   increments $Y_n \sim S_{\alpha }(1)$ are a priori unrestricted in size.

We proceed as follows.  Each random outcome   $R_n$  we evaluate in terms of modulo operation   with respect to division by   $2b$. This guarantees, that jumps of size exceeding $2b$ will be  mapped  back  (actually the remainder of the division by $2b$) to  the   interval $[0,2b)$.  In passing we note, that the jump from $0$ by an integer  multiple of $2b$, if interpreted modulo $2b$,  would always  land  at $0$.

Accordingly, if  $Y_n\geqslant0$,then  $R_n=R_{n-1} +  Y_n$  only if  $R_{n-1} +  Y_n  <b$.  On the other hand, if  $R_{n-1} + Y_n \geqslant b$, then the admissible outcome   is  $R_n=|2b- R_{n-1} - Y_n|$.

For  $Y_n<0$,  we admit  $R_n=R_{n-1}+Y_n$  only if  $R_{n-1}+Y_n >0$.   If this is not the case,  we accept the outcomes  according to: (i) if  $|R_{n-1}+Y_n|< b$, then  $R_{n}=|R_{n-1} +Y_n|$,  (ii)  if   $|R_{n-1}+Y_n|\geqslant b$, then  $R_{n}=2b-|R_{n-1}+ Y_n|$.

So defined random walk "evolution"  leads to the uniform distribution  on $[0,b]$, as anticipated. In Fig. 5 we depict results of the simulation of  $500 000$ trajectories of the pertinent reflected random walk, for  $\alpha=0.3$ at  $T=2500$. For other exemplary values of $\alpha $ the outcome (uniform distribution) is the same.

\subsection{Apprehensive stopping. Skipping the forbidden jump.}

Eq. (9) of Ref. \cite{VHS}  provides another recursion formula, used to simulate the reflection from the  bottom  barrier  ($x\geq w$) in a number of recent publications on the reflected fractional Brownian motion, \cite{M}-\cite{VHS2}. We stress that there is no unique, universally accepted reflection form the barrier scenario. In contrast to  (25) the  original  recursion formula of the  random walk:
  \begin{eqnarray}
&& x_{n+1} = x_n + \xi _n\, ;  \, \,  \,   \, x_n + \xi _n  \geq w
\nonumber \\
&& {x_{n+1} = x_n\, ;  \, \, \,  \,   x_n + \xi _n  < w }.
\end{eqnarray}
   defines an “inelastic” wall at which the  there is no move (jump) at all  from the achieved "location" $x_n$,  if the step would  ultimately  take it into the forbidden region $x < w$. The barrier is never "overshot", may merely be reached.

The above scenario can be readily adopted to the L\'{e}vy walk case. With   $X(n)=X_n$ defined by (20),  we generate the reflected process $R_n$  following (30):
\be
R_{n}=R_{n-1} +  Y_n \, ; \,     0\leqslant R_{n-1}+ Y_n\leqslant b.
\ee
If the above inequality is not  satisfied, we keep  $R_n=R_{n-1}$. e.g. skip the barrier overshooting   jumps.

The statics of "hits" of $500 000$ sample paths at "time" $T=2 500$ (with a normalised time increment)  undoubtedly reveals that the asymptotic distribution is uniform in the interval $[0,50]$.  The result  is  $\alpha $-independent, and is  depicted in Fig. 6  for  $\alpha=0.3, 1.0, 1.7$.

\begin{figure}[htp]
\begin{center}
\centering
\includegraphics[width=55mm,height=55mm]{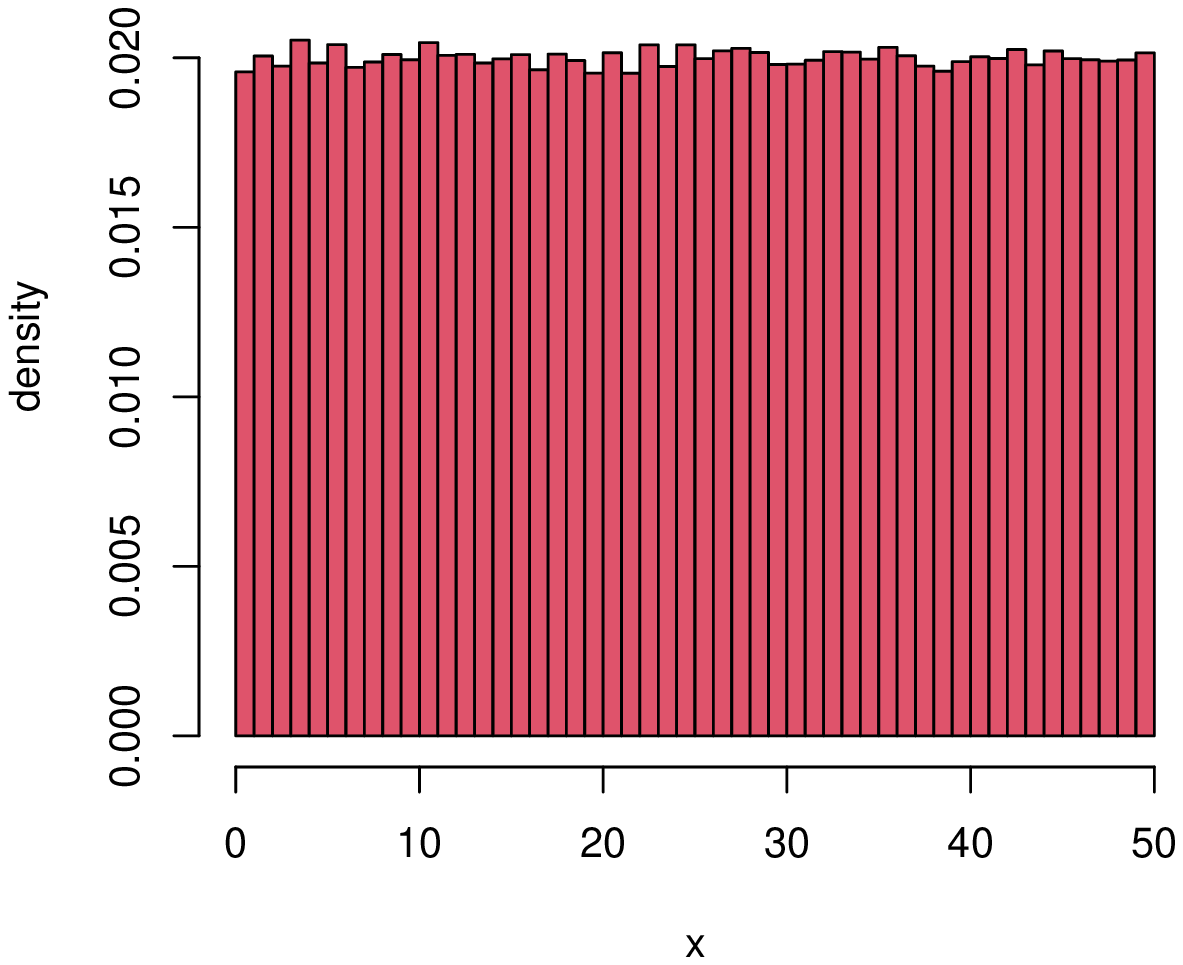}
\includegraphics[width=55mm,height=55mm]{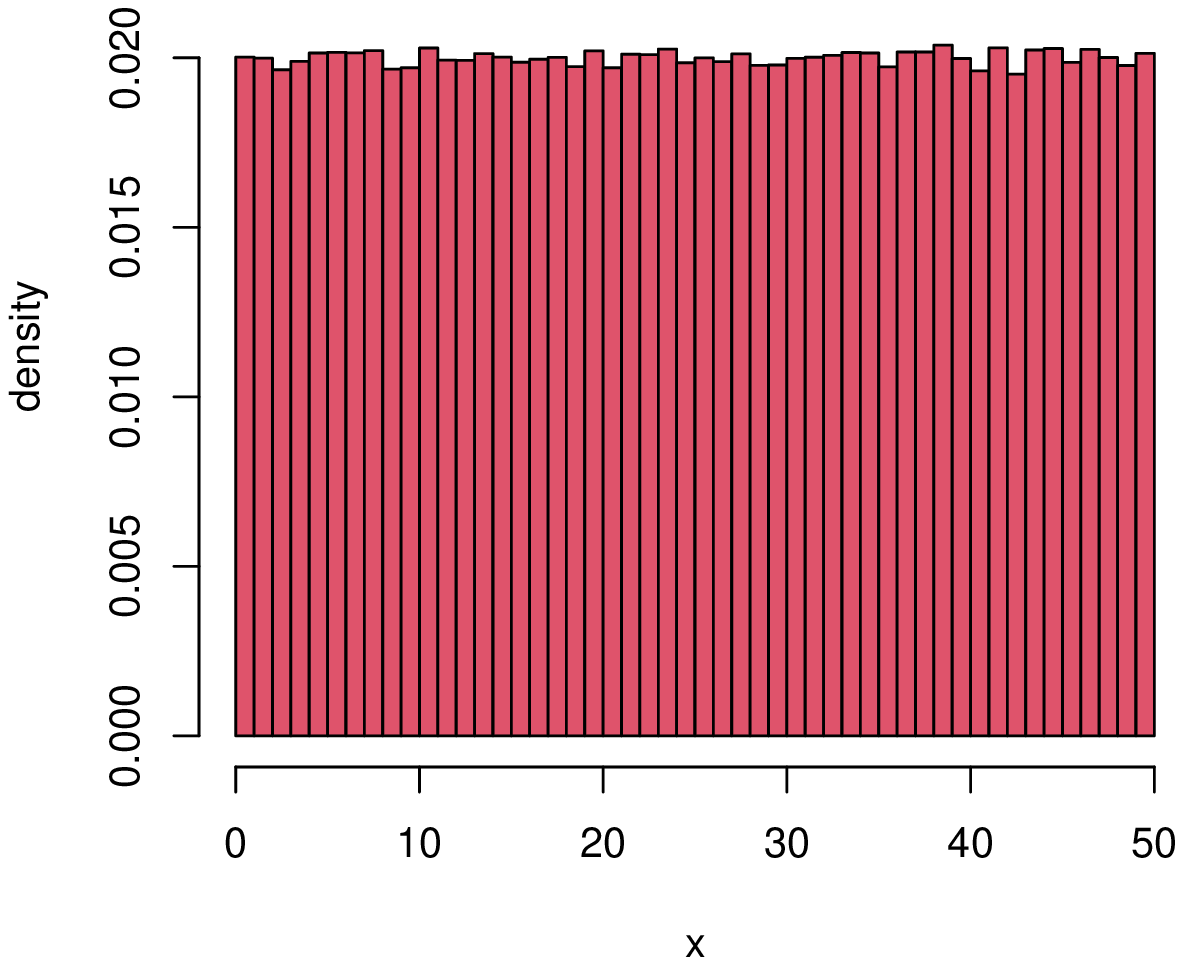}
\includegraphics[width=55mm,height=55mm]{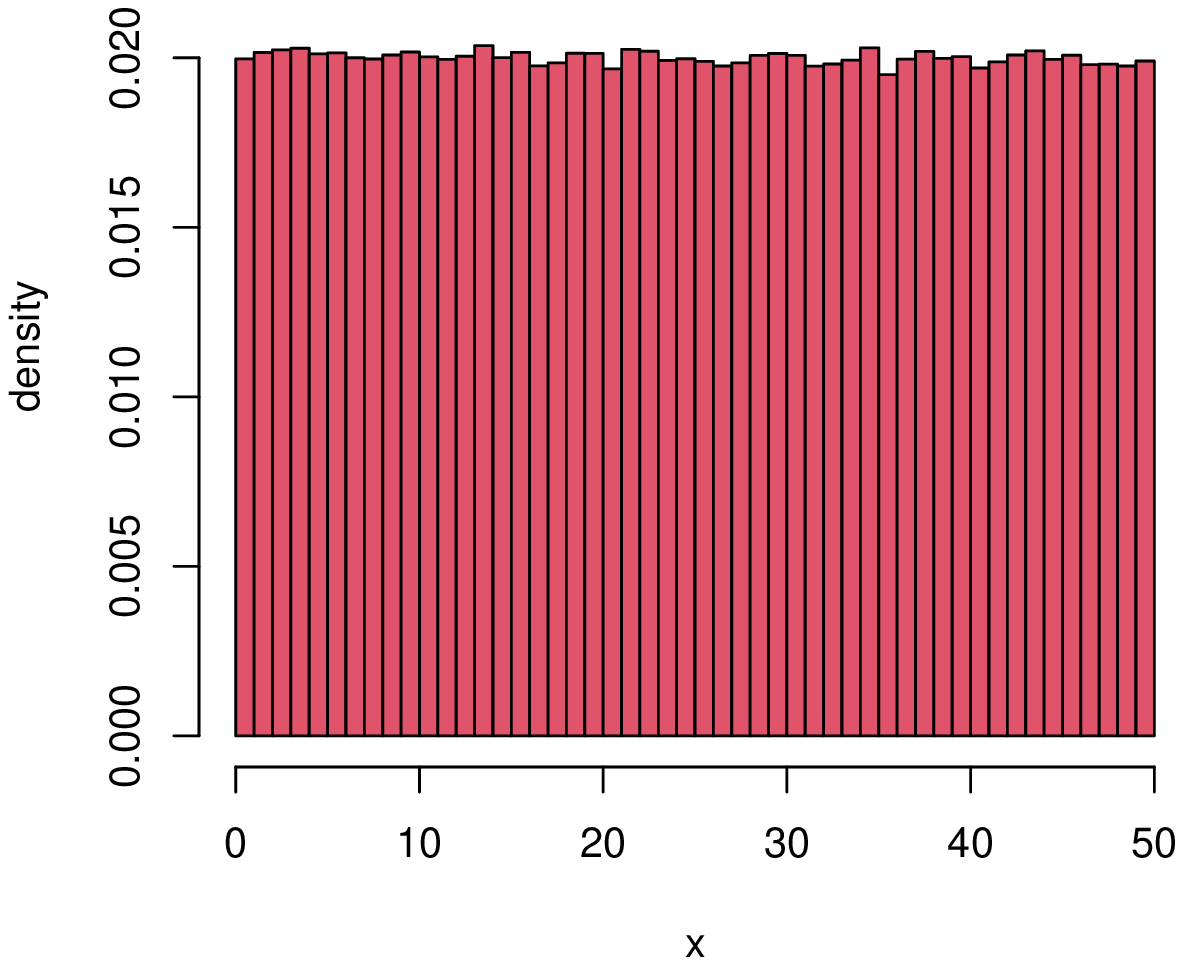}
\caption{Apprehensive stopping  in $[0,50]$. Statistics of $500 000$ trajectories for  $\alpha=0.3, 1.0, 1.7$, with $R(0) = 25$. Random walk signatures of  the uniform distribution.}
\end{center}
\end{figure}

To confirm  the distribution uniformity, we can also proceed  by evaluating the  second moment  $<R^2(t)>$   for the trajectory statistics, as a function of time.  The (expected to be) limiting value of $ <R^2>$ in  the  case of the uniform distribution on $[0,b]$  is   $b^2/3$, e.g.   about   $833.(3)$   for $b=50$.

For each trajectory started from  $R(0)=25$ , after few time steps  we identify  oscillations about $833$, which   die out with the growth of the number of trajectories, for which statistical data are gathered.  This is independent of the particular choice of $\alpha $.

In passing,we  mention that in Ref. \cite{VHS} (Fig. 1 therein)  it has been shown that for the fractional Brownian motion (FBM)  the mean value of the squared distance, in the large time asymptotic exceeds that for the uniform distribution. For the considered L\'{e}vy walk such behavior has not been  found, which supports our conjecture about the uniformity of the probability distribution in the large time limit for L\'{e}vy walks respecting the "apprehensive stopping" scenario.

\section{Nonlocal analog of the Neumann condition: Path-wise implementation.}

\subsection{Fractional heat equation with the nonlocal Neumann condition.}

In Ref. \cite{gar} we have considered seriously  a hypothesis (see e.g. \cite{bogdan,guan})  that the regional fractional Laplacian might  serve as a generator of reflected L\'{e}vy processes in the interval. This   assumption  motivated our  discussion of section V there-in, devoted to signatures of the reflecting boundaries in the spectral problem for the regional fractional  Laplacian.  By employing the Neumann basis system in the corresponding state space, we  have derived lowest eigenvalues and eigenfunctions, with the clear outcome  that  the  ground state function  is constant and  corresponds  to the  eigenvalue  zero.  This observation stays in an obvious conflict with the   formula (9), which has   been attributed to the L\'{e}vy process in the interval as well, see e.g. also \cite{gar,zaba,zaba1}.

We have learned in the previous section that some of the path-wise reflection recipes may in principle lead to uniform probability distributions in the interval, at variance with the singular $\alpha $-harmonic function shape of Eq. (9). On the other hand, we have identified (9) as the best fit approximation of the Skorohod random walk, which provides a well defined process with reflections form the interval endpoints.

In the present section, we shall outline rudiments  of another "reflecting" framework  for L\'{e}vy processes in the interval,  with the   fractional  heat (Fokker-Planck) dynamics leading asymptotically to the uniform distribution. Its  major ingredient  is the so-called nonlocal Neumann condition \cite{DRV,ros,daoud,Abatangelo}. We note that there are other Neumann condition   proposals in existence,  \cite{guan,guan1} and  \cite{bogdan},  but  transparent probabilistic   pictures, amenable to a computer-assisted (path-wise)   verification, appear to be lacking.

 The nonolocal Neumann condition of Ref. \cite{DRV}  allows to bypass these limitations.  We provide a brief resume of main results of Ref. \cite{DRV}, which is  free of (unnecessary here) technical details.

Let us come back to  an integral definition of the fractional Laplacian  (13),  while reintroducing under the integral sign  the numerator $f(x)- f(y)$ instead of $f(x)$ alone.  We extend the domain for $f(x)$ to the whole real axis and remove the exterior Dirichlet restriction upon $f(x)$  from the discussion.  This condition is removed as well from the identity (12), so that   $(-\Delta )^{\alpha /2}f(x)= h(x)$. As a matter of principle, we  may extend the   reasoning to the time-dependent problem, with $f(x)\equiv f(x,t)$, $h(x)\equiv h(x,t)$, while  assuming  an initial condition  $f(x,0) = f_0(x)$. Accordingly, by setting  $h(x,t)= - \partial _t f(x,t)$, we arrive at the  fractional Fokker-Planck type equation in $R$
\be
\partial _t f(x,t) = - (-\Delta )^{\alpha /2}f(x,t).
\ee
At the moment this equation is unrestricted by any domain requirements, and the fractional Laplacian has a standard (Cauchy principal value)  integral realization (7), (8), c.f. also  (13).

A nonlocal analogue of the classical Neumann
condition $\partial f(x)   = 0$, normally imposed   at  the boundary  $\partial D$  of the set $D$,   for L\'{e}vy processes   consists in the nonlocal prescription, \cite{DRV}:

\be
   {\cal{N}}_{\alpha }f(x)  = {\cal{A}}_{\alpha }\,
     \int_{ D} {\frac{f(x) -f(y)}{|x-y|^{\alpha +1}}}\, dy   =0,
   \ee
valid for all $x \in R\setminus \overline{D}$.  We note that if  the  integral  is restricted to $D=(0,b)$,  the Neumann condition takes the  value zero for all $x \in (b, \infty )$.  An insight into the  behavior sharply at the   the boundary point $b$,  needs  a careful execution of  limiting procedures  from the interior of $[b, \infty )$ toward $b$, \cite{DRV}.   The validity of (33) extends to the time-dependent regime  $f(x) \to f(x,t)$ as well.

Eqs. (32) and (33),   together with the initial data $f_0(x)$, constitute a heat equation with homogeneous Neumann conditions, according to Ref. \cite{DRV}.  The system, although defined on $R$ has a number of interesting properties related to the  opens set $D$, like e.g. "mass" (probability or initial normalization) conservation  inside $D$, and convergence to a constant (uniform distribution) as $t\to \infty $.  Moreover, the spectral problem for the fractional Laplacian with the boundary condition (33) has a solution such that  $(-\Delta )^{\alpha /2}u_i(x)= \lambda _i u_i(x)$ for any $x \in D$ and ${\cal{N}}_{\alpha } u_i(x)=0$ for any $x\in R \setminus \overline{D}$. The eigenvalues  are nonnegative, and  the bottom one equals zero. The eigenfunctions, if restricted to $D$, form a complete orthogonal  system in $L^2(D)$.

There is a transparent probabilistic interpretation behind the formal setting (32), (33). Namely, if $u(x,t)$ is a probability density function of a random process inside $D$, any exit beyond $D$ is {\it immediately}  followed
by a return to $D$. The way, the process comes back to $D$ according to the {\it   randomized wrapping} rule:  an  immediate return from the   "overshot" destination  $x\in R\setminus \overline{D}$  is random and gets
realised with  the return probability of jumping  from $x$ to  any    $y\in D$, which is proportional to $|x-y|^{-1 -\alpha }$.  (We recall that for the unrestricted L\'{e}vy process,a jump  from $x\in R$  to any other point $y\in R$ is realised with  the probability of jumping  being proportional to the invoked   $|x-y|^{-1-\alpha }$).

We have coined the term {\it randomized wrapping} to set a correspondence with the wrapping scenario of Section IV.B,   in which the return to $D$ is realised in the single run: start from $D$, overshoot the barrier, immediately return back to $D$.
 This tells  us  what might  mean the  "immediate return"  in the nonlocal Neumann problem.  A mirror reflection is another  example of such (albeit non-random) "immediate return".

\subsection{Instantaneous randomized wrapping.}

We have not found in the literature any explicit path-wise analysis of the above reflection scenario, hence we shall spend a while on its somewhat detailed discussion.

For  a symmetric  $\alpha$-stable  L\'{e}vy  process  $\{X(t),t\geqslant 0\}$,  its random walk version  $X_n$ is generated according to (20). The reflection scenario, we attempt to visualise by following the heuristics of Ref. \cite{DRV}, appears to be a randomised version of the previously discussed wrapping scenario.
Namely, we assume that the trajectory never actually leaves the interval $[0,b]$. All exits (potential overshooting the barier) are   {\it virtual}. The starting point is  $R(0)=b/2$.

We proceed as follows:\\
a) If  $0\leqslant R_{n-1}+ Y_n \leqslant b$ then we accept the jump $R_{n}=R_{n-1}+ Y_n$.\\
b) If  the sampled jump would be long enough to  overshoot $b$, reaching  a destination $y>b$, then the  immediate return (jump) is executed from the  coordinate $y$  to certain $x\in D=[0,b]$, with a  conditional  probability proportional to  $|x-y|^{-1-\alpha}$.

 We note that  the virtual exit $y$ from $[0,b]$ is followed by an immediate return  to a certain $x\in [0,b]$ in the single  run  (e.g.  uninterrupted  jump). The  return is {\it  immediate}  in analogy with the execution of  overshooting jumps in the wrapping scenario of subsection IV.B. Therefore, we call the current scenario  the {\it randomized wrapping} about the barrier.\\

The   probability density of  jumps from a virtual point $y\in R\setminus \overline{D} $,  to any $x \in D$.  we  denote   $\rho _y(x)=C|x-y|^{-1-\alpha}$.      We need to evaluate  the $ L^1([0,b])$ normalizing coefficient $C$. This must be done separately for the upper and lower barriers.

   Let $y>b$,  then
\be
C\int\limits_0^b |x-y|^{-1-\alpha}\,dx= \frac{C}{\alpha}\frac{y^\alpha-(y-b)^\alpha}{[y(y-b)]^\alpha}=1 \Longrightarrow    C= C_b=\frac{\alpha[y(y-b)]^\alpha}{y^\alpha-(y-b)^\alpha}.
\ee
An analogous evaluation for  $y<0$  gives rise to $C=C_0$  equal
\be
C =C_0 =  \frac{\alpha[y(y-b)]^\alpha}{(b-y)^\alpha-(-y)^\alpha}.
\ee
The respective probability densities $\rho _{y>b}(x)$ and $\rho _{y<0}(x)$ directly  follow.

Given the probability density  $\rho _{y}(x)$   of return   to $D=[0,b]$ from  any  $y\in R\setminus \overline{D}$,   we need to implement  a fully fledged randomisation of  return points $x\in D$, for each  virtual $y$ separately, while accounting for the barrier (bottom or upper) location in $R$.
To this end, we invoke  the inverse cumulative distribution function (ICDF) method, which  is a  widely recognised procedure allowing to generate random samples, that are   consistent with any   prescribed  probability distribution, \cite{dev}   Chap. II.

Given  $y>b$, to deduce the   random return coordinate $x \in D$  ("reflection" point  for a jump turned back at $y>b$ ), let us denote  $p$  a  value  of the  random variable $U$ sampled  from the uniform distribution on $[0,1]$.  We require  that each sampled  $p\sim U(0,1)$ is {\it  uniquely }  assigned to the return point $x \in D$. This we  secure  in terms of the  cumulative probability distribution evaluated up to the point $x$:
\be
\int\limits_0^x \rho_{y>b}(z)\,dz = p.
\ee
  To infer the return coordinate $x\in [0,b]$  of the completed jump (overshooting and random return) of the sample  trajectory, we must employ  the inverse function  $F^{-1}_X(p) =x$, which  uniquely identifies $x\in D$, given $p\in [0,1]$. This   randomization procedure refers to each  wrapping  point $y>b$ separately.

We have
\be
C\int\limits_0^x |z-y|^{-1-\alpha}\,dz = C\int\limits_0^x (y-z)^{-1-\alpha}\,dz = \frac{C}{\alpha}\left[(y-x)^{-\alpha}-y^{-\alpha}\right]= p.
\ee
After inserting $C=C_b$ of Eq. (35), we ultimately get
\be
x= x_b= y-y\left[1-p+p\left(\frac{y}{y-b}\right)^\alpha\right]^{-1/\alpha}.
\ee
The random sampling of $p$ has been uniquely transferred to the randomness of $x$-outcomes. Indeed,  for a uniformly distributed  $p\in [0,1]$, the probability distribution of  $X$ with values $x$ given by  Eq. (38) has a probability density function $\rho _{y>b}(x)$.

Analogously, for  $y<0$  the jump return destinations  $x\in[0,b]$, derives as
\be
x=  x_0= y-y\left[1-p+p\left(\frac{-y}{b-y}\right)^\alpha\right]^{-1/\alpha}.
\ee
Random outcomes $x_0$ and $x_b$, can  actually be  obtained from a formula, encompassing both cases.
Indeed, the random  return jump location   $x= x(y)$   is given by a compact formula
\be
x= x(y)=y-y\left[1-p+p\left|\frac{y}{y-b}\right|^\alpha\right]^{-1/\alpha},
\ee
for each $y\notin[0,b]$ and each preassigned (sampling from a uniform distribution) value of $p$.  \\

With these preparations, we are  finally ready to  accomplish the visualisation   of the random wrapping reflection scenario, according to Ref. \cite{DRV}. Steps a) and b) described above  are now completed by one more step:\\
c) If   $y=R_{n-1} +Y_n$  is      beyond $[0,b]$, then the  final destination of the jump,  originating from $R_{n-1}$  (jump with wrapping return),   is given by  $R_{n}=x$, where   $x$ stands for the  random coordinates in $D$, which is uniquely related to a pre-sampled value of   $p\sim U(0,1)$, c.f. . \\

In Fig. 7 we have depicted a statistics of  $500 000$ trajectories  of times span $T=2 500$, with the normalised time step, choosing $b=50$ and following the  reflection scenario a) to c), for  $\alpha=0.3, 1, 1.7$.

\begin{figure}[htp]
\begin{center}
\centering
\includegraphics[width=55mm,height=55mm]{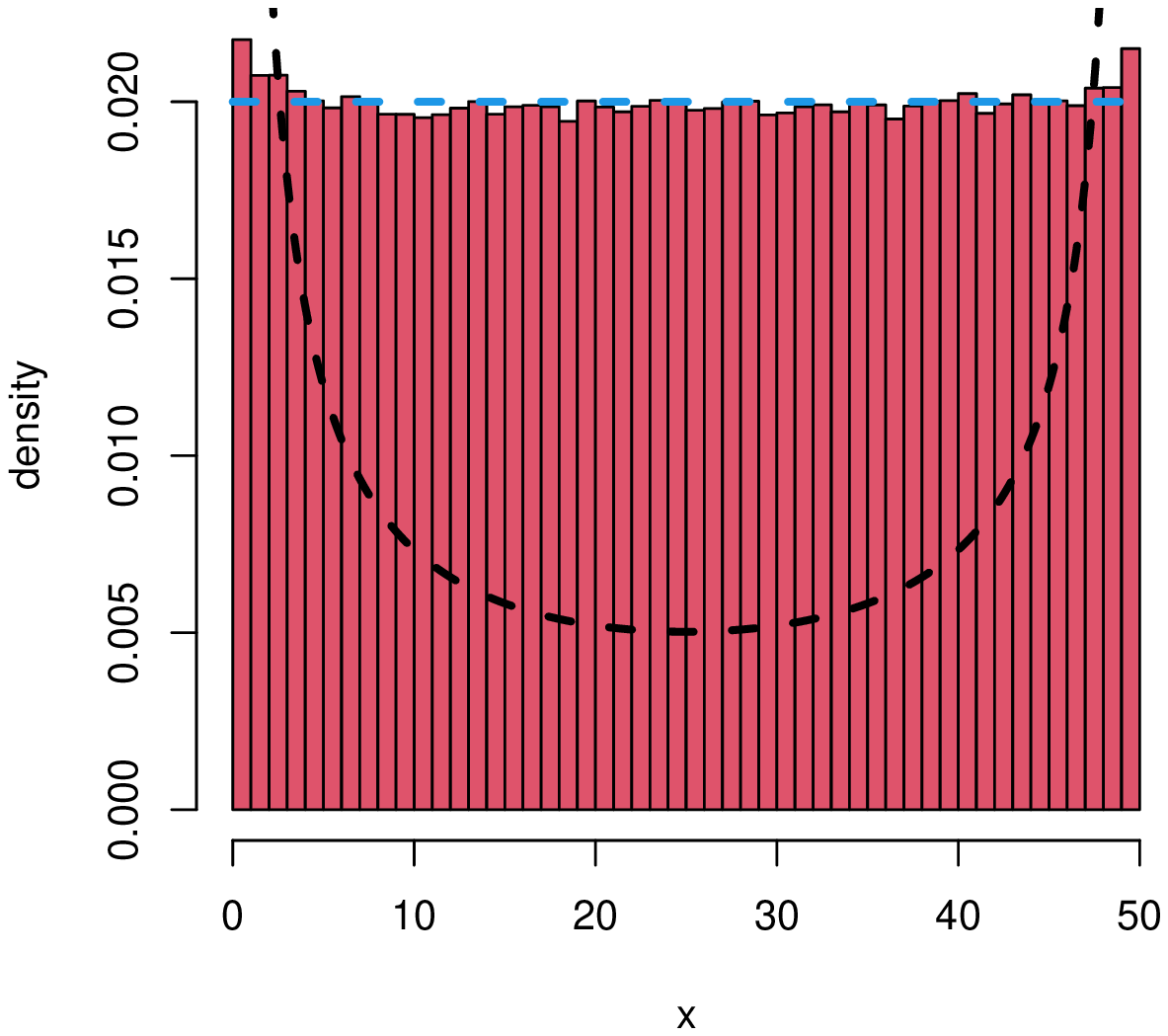}
\includegraphics[width=55mm,height=55mm]{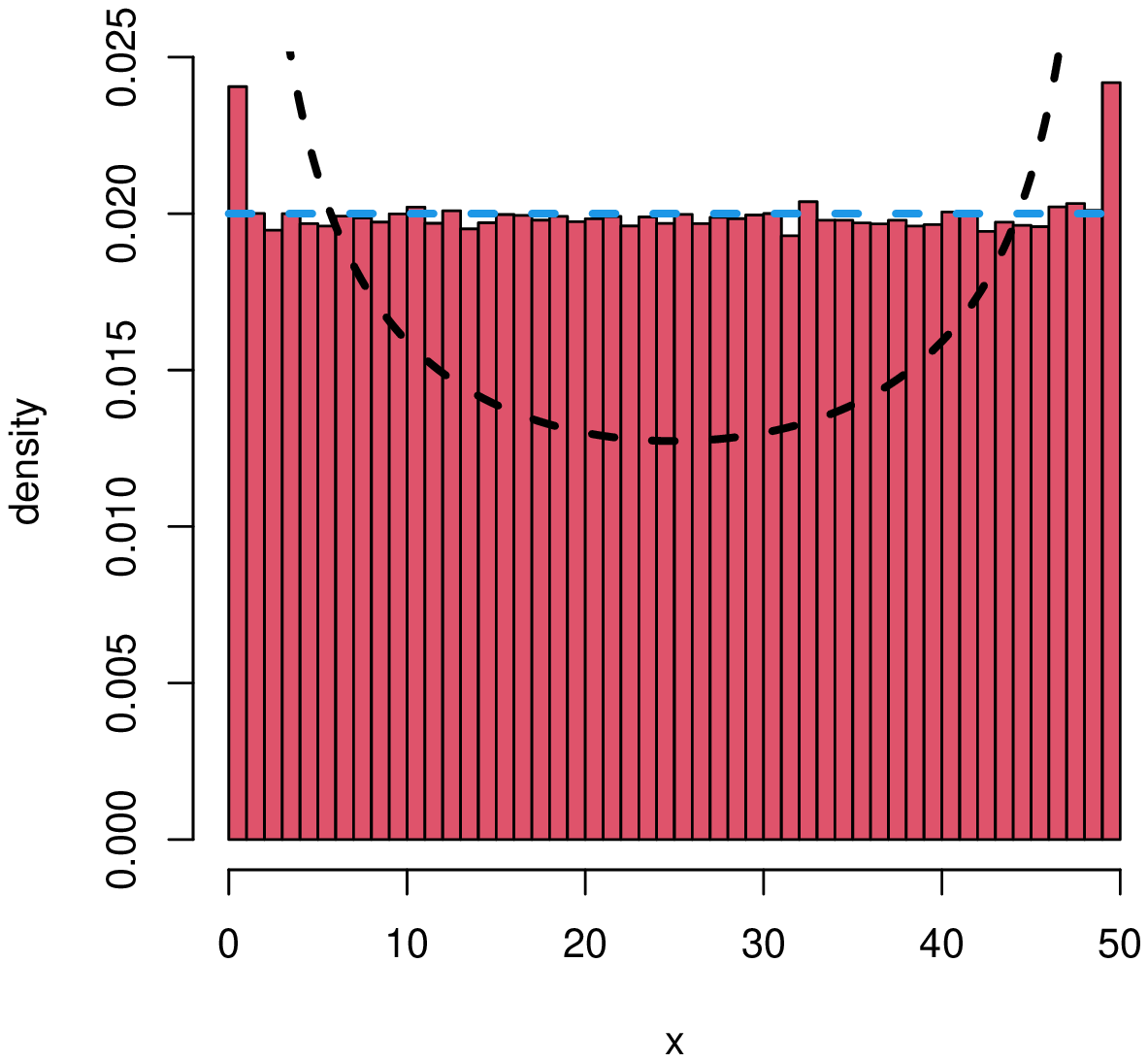}
\includegraphics[width=55mm,height=55mm]{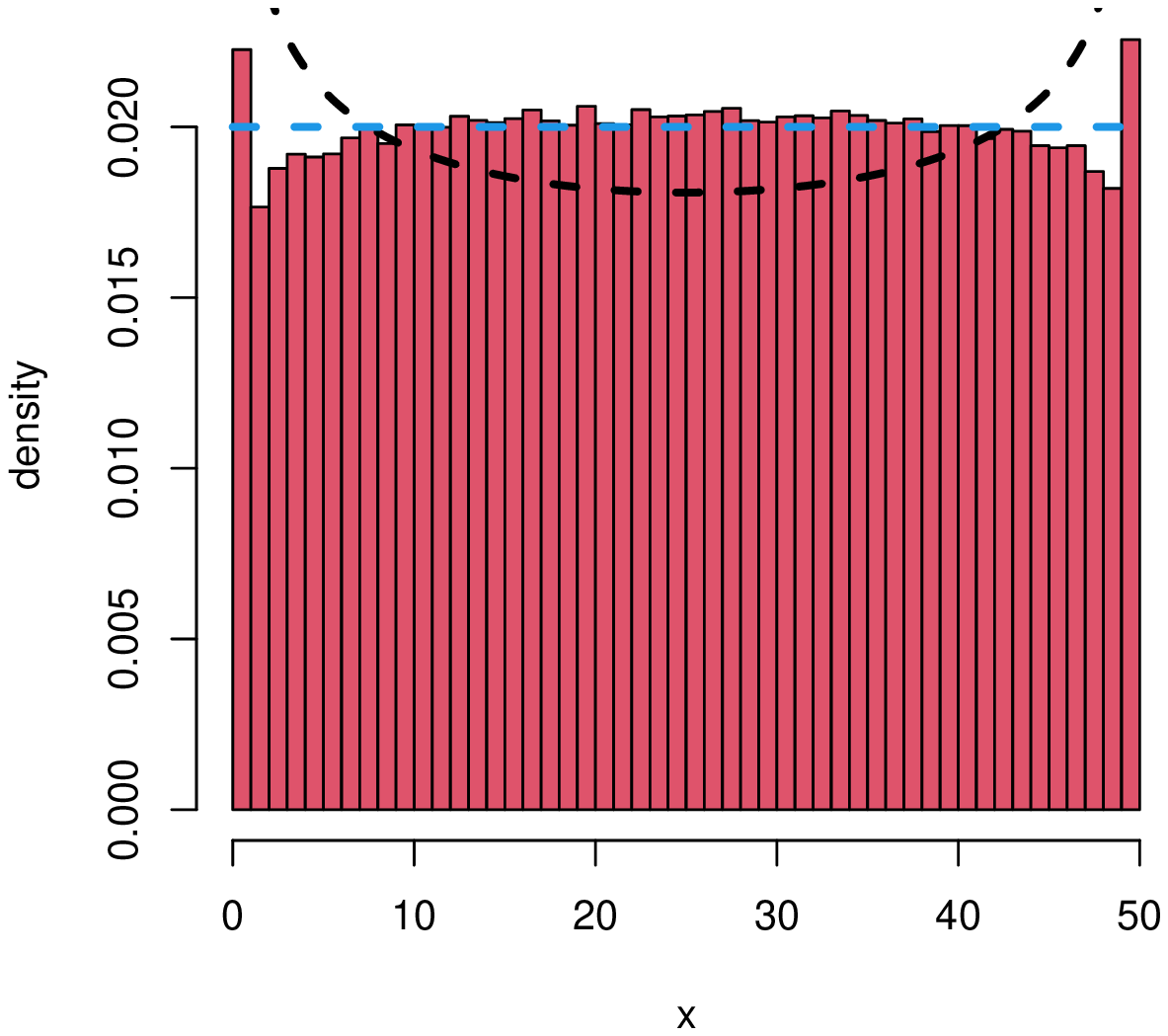}
\caption{Randomised wrapping.  Statistics of  $500 000$ trajectories for  $\alpha=0.3, 1, 1.7$. The starting point is $v=25$. The black   dashed curve  comes from the formula (10) and is introduced for reference. The green line indicates an approximating uniform distribution  $U(0,50)$.  This result is consistent with the  asymptotic behavior of the solution of the fractional heat equation with the nonlocal Neumann condition, Eqs. (32), (33).}
\end{center}
\end{figure}

For  $\alpha=0.3$ (left panel in Fig.7) the distribution is fapp (for all practical purposes) constant on $[0,50]$, except for a close vicinity of the interval endpoints, where frequency histograms slightly  grow with a diminishing distance. This behavior can be interpreted  as follows. For small  $\alpha$ long  jumps are relatively frequent, so that quite often their virtual destinations are beyond $[0,b]$.  The randomised wrapping and return of the jump to $[0,b]$ is ruled  by the probability density $\rho(x| y)$, which appears to favor final destination close to the endpoints, against these close to the central part of the interval.

On the other hand, for  $\alpha=1.7$ (right panel in Fig. 7) frequency of long jumps is significantly reduced  and statistically important virtual overshoots of the barriers are dominated by  these originating from points close to  barrier in the interior of $[0,b]$. The random returns do not seem to compensate the probability loss near the barriers, and contribute to  the  remaining part of $[0,b]$. The case of  $\alpha=1$  appears to be transitional in this respect, and  shows a mutual compensation of the outlined before ($\alpha =0.3$ vs $\alpha =1.7$) probability redistribution tendencies.

Anyway, the  theory of Ref. \cite{DRV} says that in the long time asymptotic, the uniform distribution should be ultimately reached  for all $\alpha  \in (0,2)$.

\subsection{Alternative  model. Delayed (separate run) randomised return.}

Simply, out of curiosity, let us  consider a modification of the previous scenario, which might have  some realistic physical appeal.  Let us admit the overshooting is a realistic event, and the path exits form $D$ to the exterior point $y$.  To divert the trajectory back to $D$, we  presume to  need a separate (randomized) jumping event, with the  jump length ruled by the probability density   $\rho _y(x)$ of the previous subsection.

 All formulas of the previous subsection retain their validity, except that $y$ is not a virtual point, but a real destination of the overshooting jump through any barrier of $[0,b]$.  Consequently,  the exit point  $y$, in the next time step  becomes a starting point for the independent jump (with length randomized according to  $\rho _y(x)$),  sending   the trajctory back to $[0,b]$.

Statistical data for  $500 000$ trajectories, generated with a normalised time step, and $b=50$, have been collected accepting the {\it delayed } reflection scenario, for $\alpha=0.3, 1, 1.7$. These are depicted in Fig. 8.

\begin{figure}[htp]
\begin{center}
\centering
\includegraphics[width=55mm,height=55mm]{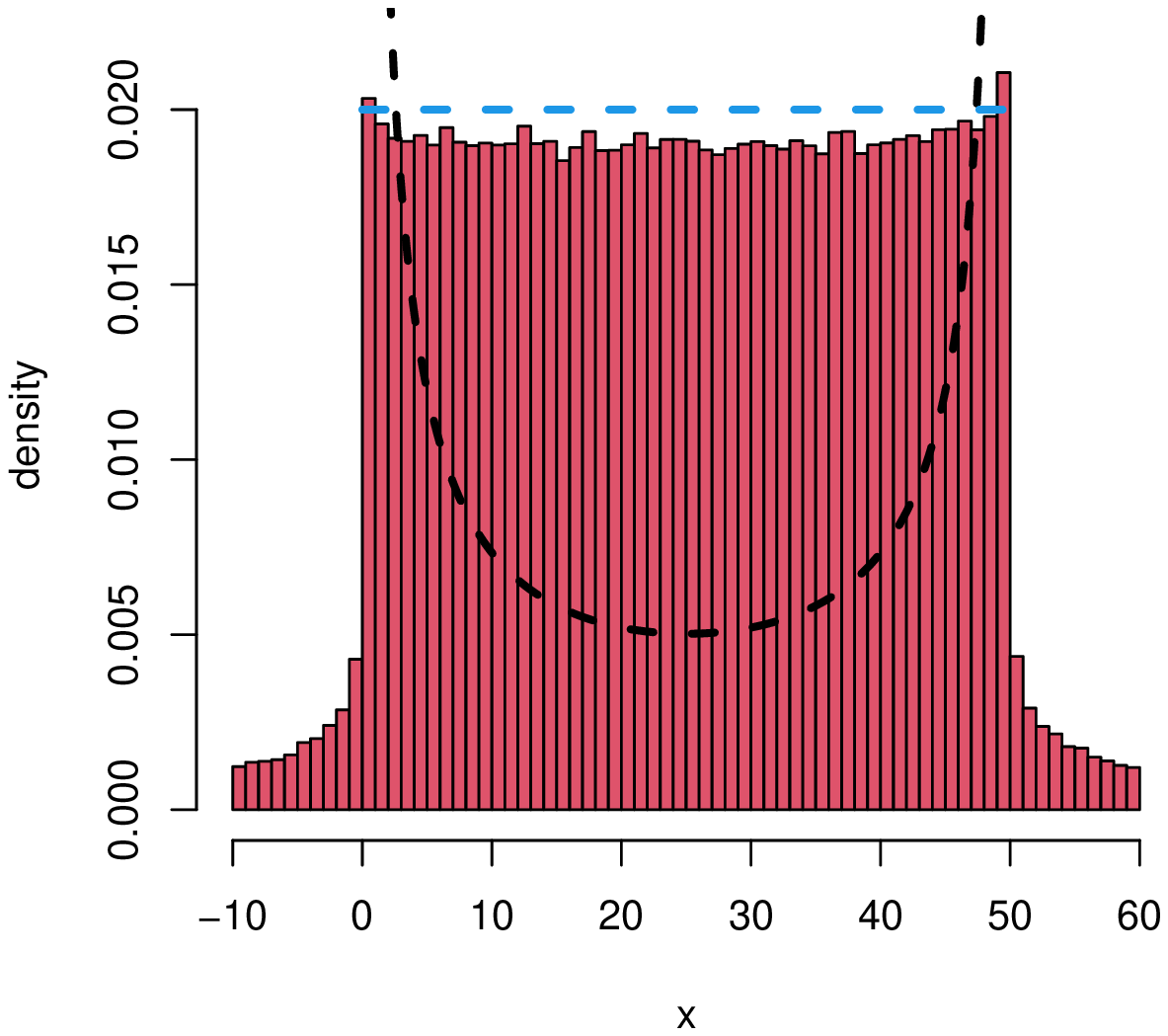}
\includegraphics[width=55mm,height=55mm]{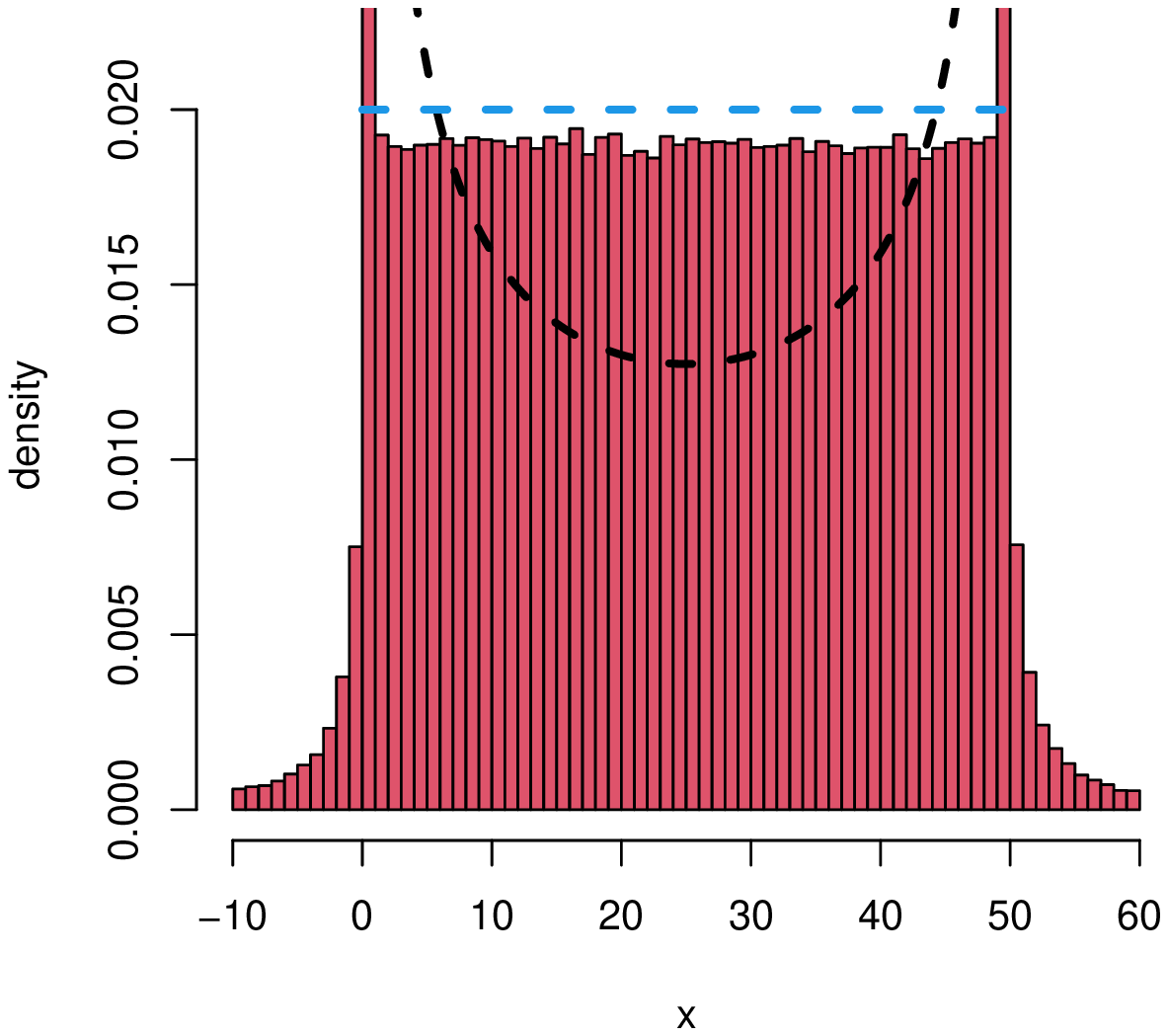}
\includegraphics[width=55mm,height=55mm]{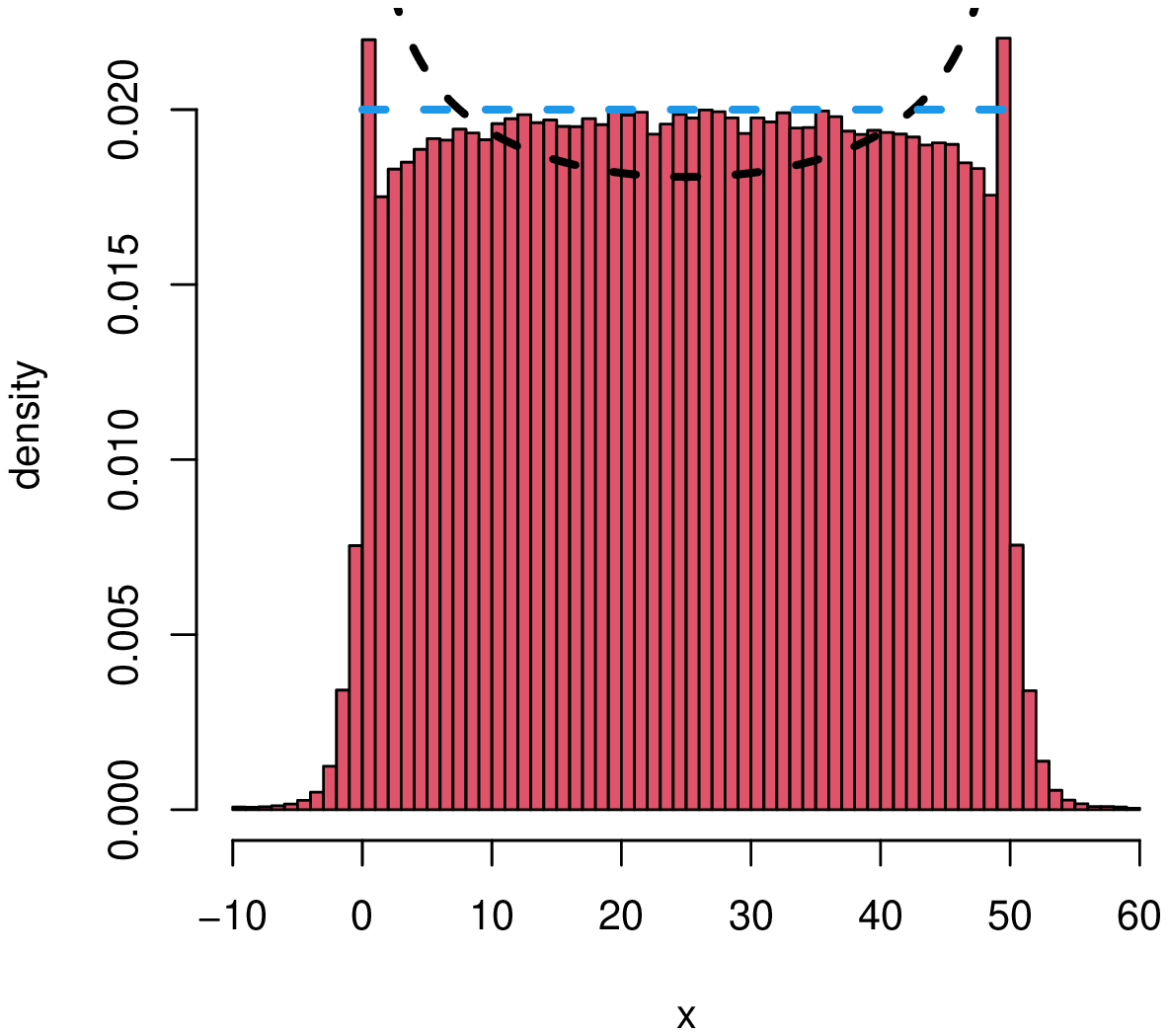}
\caption{Delayed randomized return.   Statics of  $500 000$ trajectories for  $\alpha=0.3, 1, 1.7$. Trajectories are started at $x=25$.  Black  dashed  curve depicts the reference curve (9)  for $b=50$, while the green dashed  line  refers to the uniform distribution $U(0,50)$.}
\end{center}
\end{figure}

Qualitatively, the statistic (histograms)  of hits at time $2 500$, are close to these produced   in the randomized wrapping reflection of the Subsection V.B.  Beacuse we allow trajectiories to leave $D$, with return in the next time step, there are cleraly visible distribution "tails", beyond $[0,50]$.  "Heavy" (long) tails effects are clearly  displayed, specifically for low value of  $\alpha  =0.3$, and their contribution gets minimized with the growth of $\alpha $.

If to rescale  the time step  form $1$ to $1/8$, we   uncover a clear similarity (except for remnant distribution  tails  in close outside vicinity of the barriers) to the randomized wrappin results of Section V.B.   The corresponding statistical data (histograms) for  $500 000$ trajectories  are depicted in Fig. 9.

\begin{figure}[htp]
\begin{center}
\centering
\includegraphics[width=55mm,height=55mm]{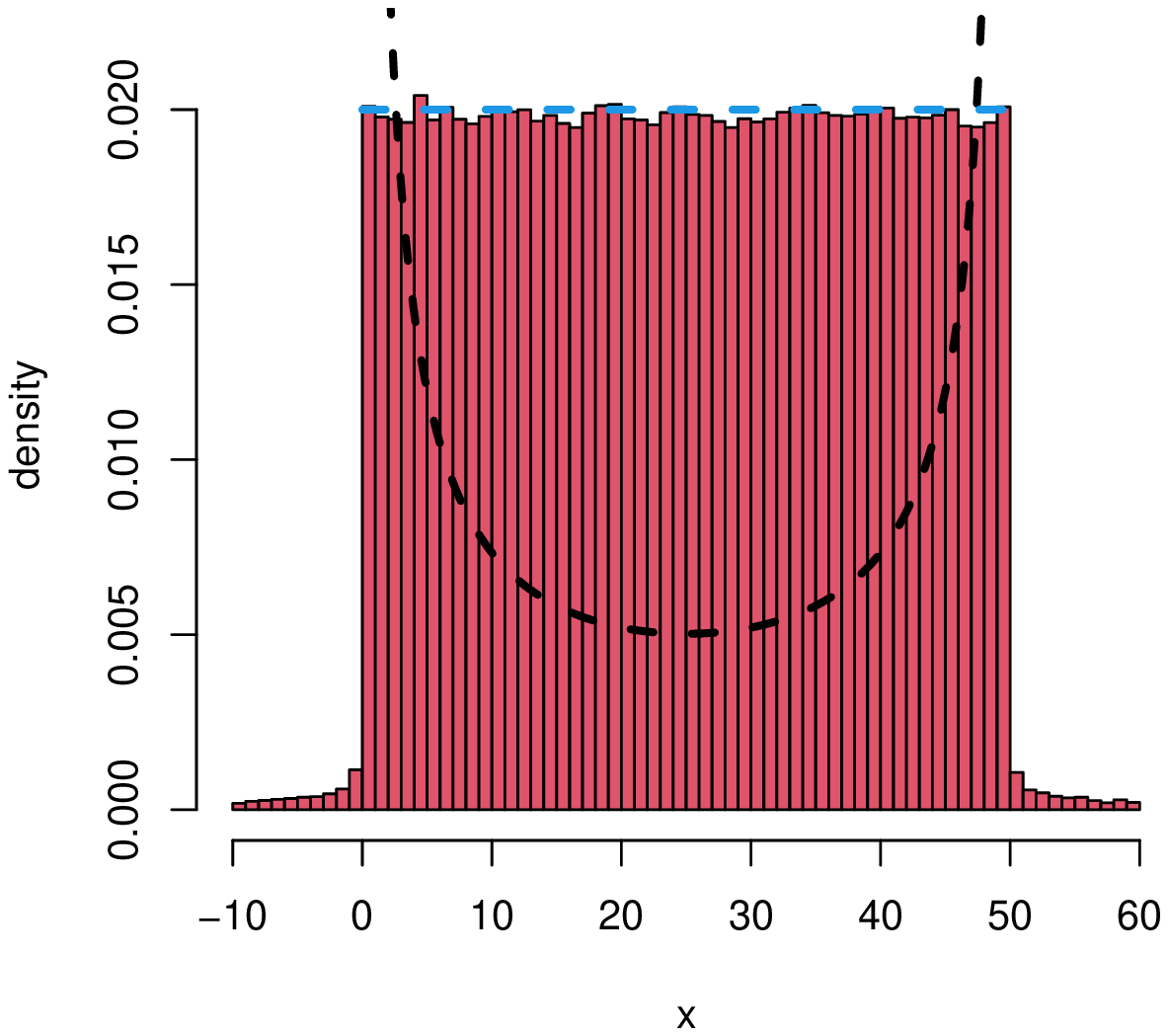}
\includegraphics[width=55mm,height=55mm]{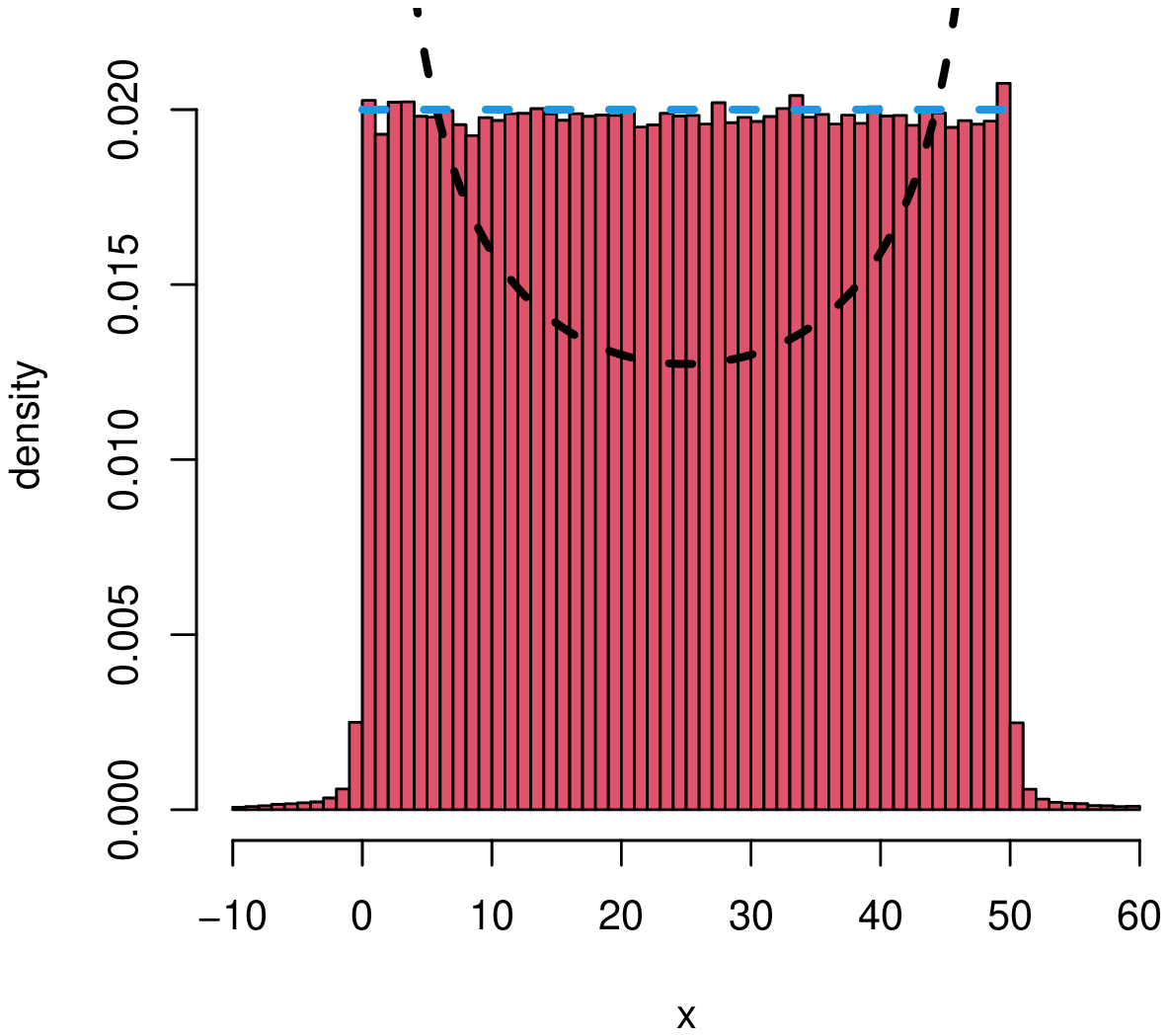}
\includegraphics[width=55mm,height=55mm]{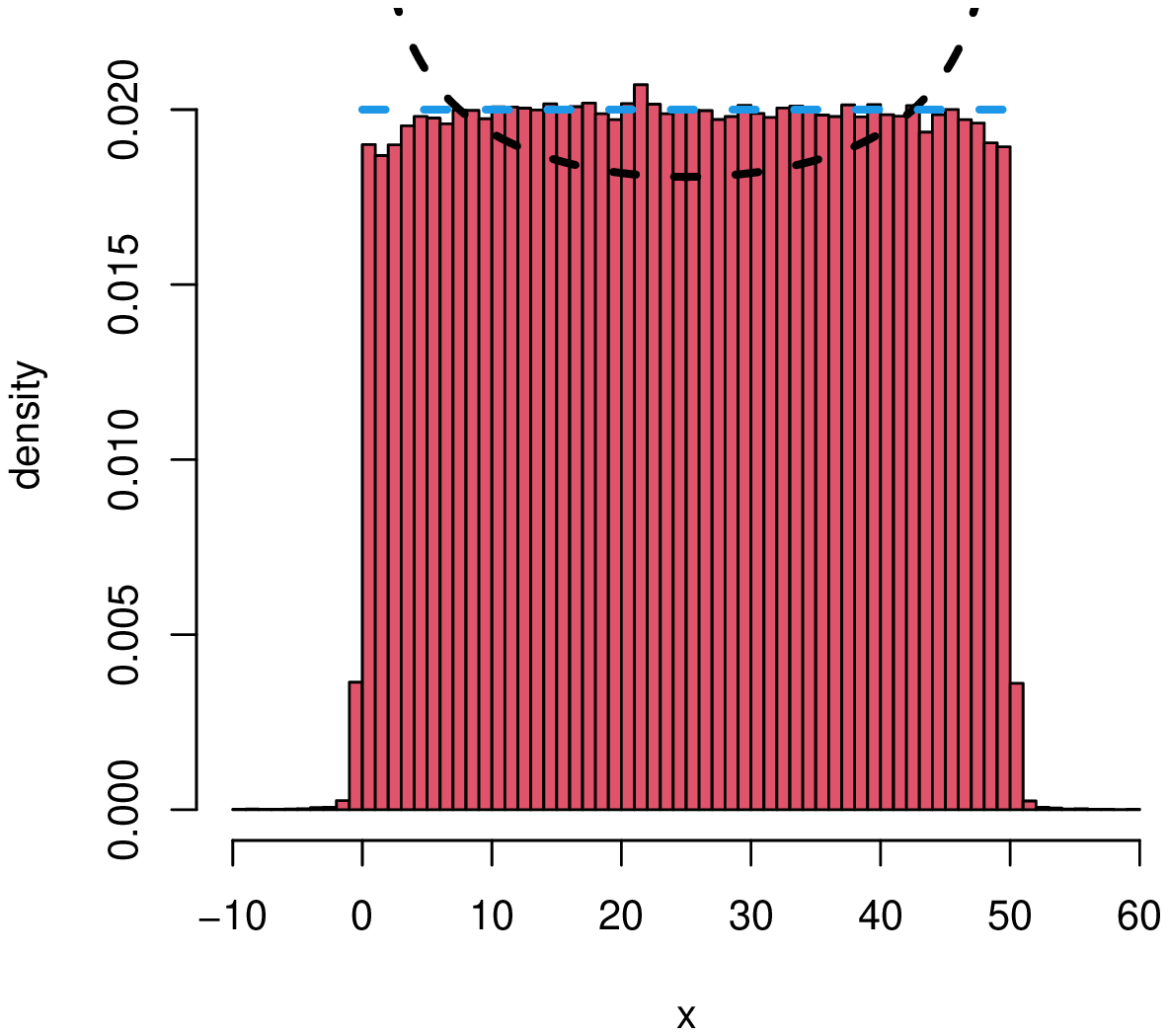}
\caption{Delayed return. Statistics (histograms) of $500 000$ trajectories for  $\alpha=0.3, 1, 1.7$, with the time step rescaled to $1/8$. Trajectories start from   $x=25$. The black curve refers to Eq. (9), while the green dashed line depicts  the uniform distribution $U(0,50)$.}
\end{center}
\end{figure}

Accordingly, with the small time step, statistical data for  the L\'{e}vy walk with the  randomized wrapping at barrier, become fapp (for all practical purposes) indistinguishable from  these obtained in the delayed return scenario. Minor long tail remnants beyond the barriers, for a large number of time steps  and the "small time increment"  discretization,  practically   may be absorbed in the "discretisation inaccuracy" estimates.

\section{Conclusions.}

Essentially new results in   our  path-wise analysis of L\'{e}vy random walks (interpreted as time-discretizations of  regular $\alpha $-stable L\'{e}vy processes), are contained in Sections III and V. Arguments of Section IV give a supplementary view upon path-wise  reflection scenarios  employed in the current physics-oriented research (in reference to both the fractional Brownian motion, \cite{M}-\cite{VHS2},   and L\'{e}vy flights proper, \cite{dybiec}-\cite{denisov}).

Instead of imposing the boundary   restrictions upon fractional motion generators (c.f. Section II), we have started from the L\'{e}vy random walk approximation of
$\alpha $-stable jump-type processes on $R$.  In section III,  an explicit random walk construction of the Skorohod reflection process in the interval has been performed. We have obtained convincing statistical data about confining properties of this walk.  The asymptotic pdf (its histogram) is satisfactorily approximated by the singular $\alpha $-harmonic function  (9) (deduced by other means in Ref. \cite{denisov}).  The approximation accuracy can be improved  while increasing the time-span of the walk and improving the discretization finesse of time-increments. To our knowledge, for the first time an explicit  functional form of the asymptotic  pdf has been associated with the Skorohod random walk.

As indicated in Section IV asymptotic pdfs  are quite sensitive to the choice of a concrete  reflection-at-the-barrier scenario. The stopping procedure of Section IV.A  implies  the singular $\alpha $-stable pdf   on the trajectory statistics  level. The  procedure  of \cite{dybiec} stays in close affinity with that scenario and in fact is a realisation of the Skorohod reflecting walk in the interval reduced by $2\epsilon $. To be more concrete, instead of  the reference interval  $[0.50]$ employed by us in Section III, one bypasses the problem that (9) is a  valid harmonic function in the interval $(0,50)$, being equal to zero in $R\setminus (0,50)$, by considering (effectively) the Skorohod random walk problem in the interval $[\epsilon , 50- \epsilon ]$, where $\epsilon = 0.001$, \cite{dybiec0}.

Other popular reflection scenarios (Sections IV.B and IV.C) induce uniform  asymptotic  pdfs in the  interval (this is consistent with spectral solutions for the regional Laplacian, \cite{gar}).

In Section V we have discussed the nonlocal Neumann condition proposal of Ref. \cite{DRV,ros,Abatangelo}, presenting an explicit construction of the related  random walk, in conjunction with the asymptotic pdf data. By theory of \cite{DRV}, the pertinent pdf should be uniform in the interval. Our statistical analysis is compatible with the result.

As an alternative reflection scenario (this is  not covered by the original paper \cite{DRV}, we have investigated the {\it delayed randomized} L\'{e}vy walk whose  sample paths are allowed to exit the confining interval, but in the next time step (that is the delay) they return back to the pertinent interval, according to the random rule of Section V.A.  The resultant asymptotic pdf  shows signatures of uniformity in the interval, with  fapp (for all practical purposes) negligible long-tail remnants in the close  outside  vicinity of the interval endpoints.

We point out that the explicit solution of the Skorohod random walk problem in the interval, allows to resolve a conundrum \cite{gar,zaba} arising in connection with derivations of  the formula (9)  in  Ref. \cite{denisov}. Namely, the  reasoning of  \cite{denisov} begins from assuming the validity of the exterior Dirichlet   boundary data for the fractional Laplacian in the (open) interval.  It is well known, that  so restricted fractional Laplacian, admits well defined strictly positive (eigenvalues)   spectral solution, with bounded  eigenfunctions, \cite{zaba0,zaba3,zaba4}  and \cite{kwasnicki}-\cite{kulczycki}.
We realize that (9) is an example of the unbounded function.

Uniform probability distributions in the interval,  can be consistently related with regional fractional Laplacians, \cite{gar, ibrahimov,ievlev}  and the nonolocal Neumann condition of Ref. \cite{DRV}. In these cases,  one may deduce  spectral solutions  with the bottom eigenvalue zero and a constant eigenfunction.

In our discussion, we have described the appearance of  two  basic  types of asymptotic pdfs - uniform and singular (c.f. (9)) -  which can be associated with the two-sided reflected L\'{e}vy process.  We do not know of any other possibilities, but a discussion of censored L\'{e}vy processes in Ref. \cite{bogdan} seems to leave some  (possibly narrow)  room for other path-wise confinement scenarios, with potential consequences for the asymptotic pdf shapes. In our opinion the term "reflection" still remains  ambiguous therein.  As well, we do not know of  any  explicit shape analysis of asymptotic pdfs for   a  family of  reflected L\'{e}vy processes, analyzed in Ref. \cite{guan,guan1}, except for the statement that regional  fractional Laplacians are appropriate motion  generators.  This however,  might refer to  uniform probability distributions in the interval, in conformity with  arguments of Ref. \cite{gar}, c.f. Section V  therein.

\end{document}